\documentclass[3p,12pt]{elsarticle}

\usepackage{amssymb,amsmath,xcolor,float}

\usepackage{hyperref}
\usepackage{cleveref}

\newcommand{\Tr}{\text{Tr}}

\newcommand{\braket}[2]{\big<#1|#2\big>}

\newcommand{\qd}{$q$-deformed\ }
\newcommand{\qdf}{$q$-deformation\ }
\newcommand{\qq}{$q$\ }

\newcommand{\ben}{\begin{eqnarray}\displaystyle}
\newcommand{\een}{\end{eqnarray}}

\newcommand{\be}{\begin{equation}}
\newcommand{\ee}{\end{equation}}

\newcommand{\bc}{\begin{center}}
\newcommand{\ec}{\end{center}}

\newcommand{\eesp}{\end{split}}
\newcommand{\bsp}{\begin{split}}

\newcommand{\Rmnum}[1]{\expandafter\@slowromancap\romannumeral #1@}

\renewcommand{\i}{\iota}		
\renewcommand{\o}{\omega}	
\newcommand{\q}{\theta}	
	
\renewcommand{\r}{\rho}		

\renewcommand{\t}{\tau}		

\newcommand{\cH}{\mathcal{H}}

\newcommand{\cZ}{\mathcal{Z}}

\newcommand{\ra}{\rightarrow}

\newcommand{\lB}{\left [}
\newcommand{\rB}{\right ]}
\newcommand{\lb}{\left (}
\newcommand{\rb}{\right )}

\newcommand{\for}{\text{for}}
\newcommand{\where}{\text{where}}
\newcommand{\with}{\text{with}}
\newcommand{\tand}{\text{and}}

\newcommand{\bensp}{\begin{eqnarray}\begin{split}}
\newcommand{\eensp}{\end{eqnarray}\end{split}}

\newcommand{\bnm}{\begin{matrix}}
\newcommand{\enm}{\end{matrix}}

\def\Xint#1{\mathchoice
{\XXint\displaystyle\textstyle{#1}}%
{\XXint\textstyle\scriptstyle{#1}}%
{\XXint\scriptstyle\scriptscriptstyle{#1}}%
{\XXint\scriptscriptstyle\scriptscriptstyle{#1}}%
\!\int}
\def\XXint#1#2#3{{\setbox0=\hbox{$#1{#2#3}{\int}$ }
\vcenter{\hbox{$#2#3$ }}\kern-.6\wd0}}

\newcommand{\ket}[1]{|#1\big>}
\newcommand{\bra}[1]{\big<#1|}

\numberwithin{equation}{section}

\makeatletter
\def\ps@pprintTitle{%
	\let\@oddhead\@empty
	\let\@evenhead\@empty
	\def\@oddfoot{}%
	\let\@evenfoot\@oddfoot}
\makeatother

\biboptions{sort&compress}
\begin{document}
	
\begin{frontmatter}
\title{A Unitary Matrix Model for $q$-deformed Plancherel Growth}
\author[1]{Suvankar Dutta}
\ead{suvankar@iiserb.ac.in}
\author[2]{Debangshu Mukherjee}
\ead{debangshu0@gmail.com}
\author[1]{Neetu}
\ead{neetuj@iiserb.ac.in}
\author[1]{Sanhita Parihar}
\ead{sanhita18@iiserb.ac.in}
\address[1]{Department of Physics Indian Institute of Science Education and Research  Bhopal,\\ Bhopal bypass, Bhopal 462066, India}
\address[2]{Indian Institute of Science Education and Research  Thiruvananthapuram,\\ Maruthamala PO, Vithura, Thiruvananthapuram - 695551, Kerala, India.}

\begin{abstract}
In this paper we construct a unitary matrix model that captures the asymptotic growth of Young diagrams under \qd Plancherel measure. The matrix model is a \qq analog of Gross-Witten-Wadia (GWW) matrix model. In the large $N$ limit the model exhibits a third order phase transition between no-gap and gapped phases, which is a \qd version of the GWW phase transition. We show that the no-gap phase of this matrix model captures the asymptotic growth of Young diagrams equipped with \qd Plancherel measure. The no-gap solutions also satisfies a differential equation which is the $q$-analogue of the automodel equation. We further provide a droplet description for these growing Young diagrams. Quantising these droplets we identify the Young diagrams with coherent states in the Hilbert space. We also elaborate the connection between moments of Young diagrams and the infinite number of commuting Hamiltonians obtained from the large $N$ droplets and explicitly compute the moments for asymptotic Young diagrams.
\end{abstract}
		
		
	\end{frontmatter}


\tableofcontents

\section{Introduction}
\label{sec:intro}

The powerful techniques of random matrix models have been applied to numerous fields in physics and mathematics. Quantum gravity in 2D, knot theory, integrable systems, number theory, topological string theory, algebraic geometry are a few from the list of many. In this paper, we study the asymptotic growth of Young diagrams, equipped with \qq analog of Plancherel measure and show that one can write a unitary matrix model (UMM) to capture such growth processes. The matrix model is a \qq analog of Gross-Witten-Wadia (GWW) matrix model. The model exhibits a third order phase transition which is \qd version of GWW phase transition. We observe that the no-gap phase of this UMM captures the \qd Plancherel growth of Young diagrams. This is an extension of our earlier work on asymptotic growth of Young diagrams under ordinary Plancherel measure \cite{Chattopadhyay:2019pkl}.

\begin{figure}[h]
	\begin{center}
		\includegraphics[scale=0.4]{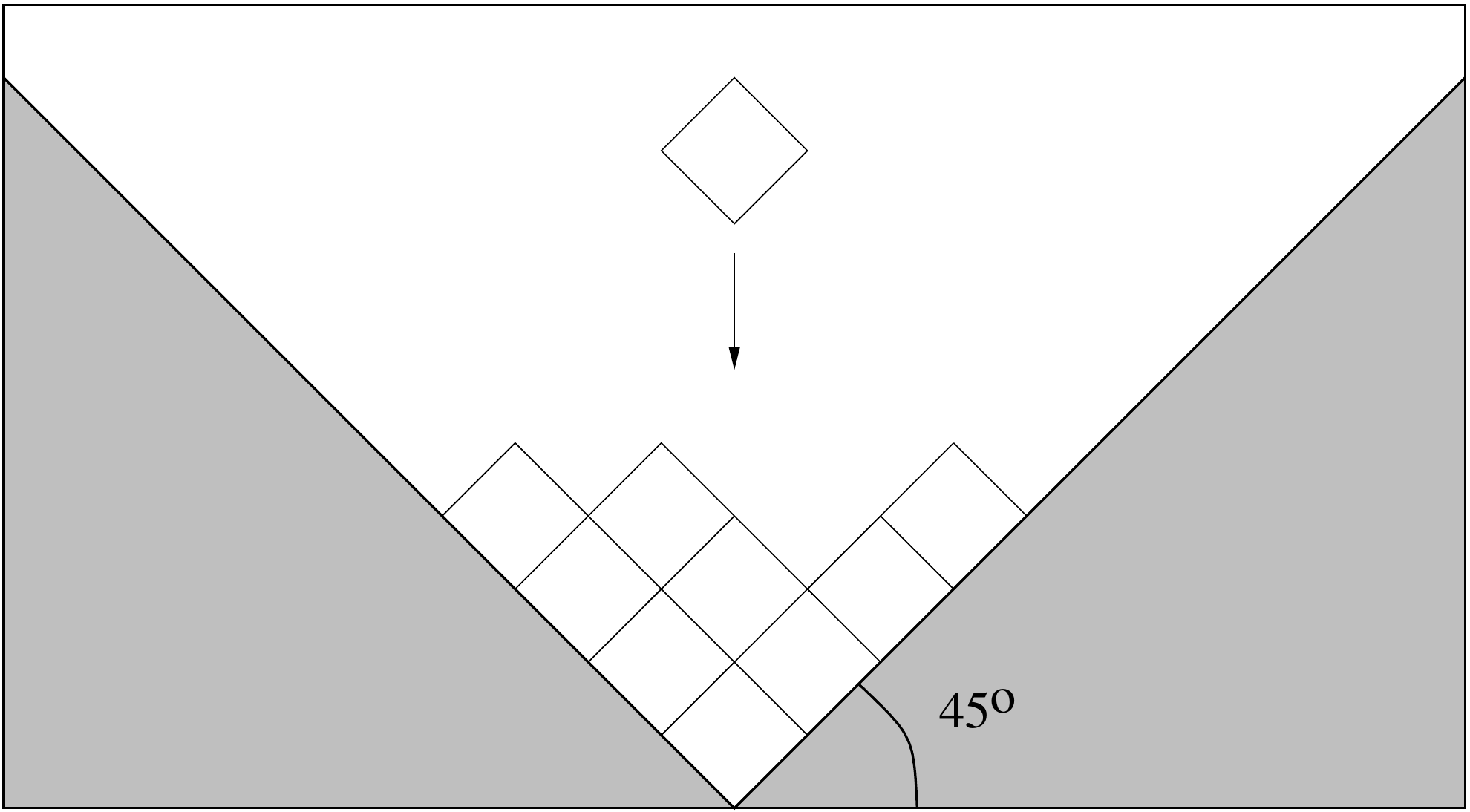}
		\caption{2D Crystal growth and growth of rotated Young diagrams. A crystal unit is compared with a single box in the Young diagram. Single crystals are falling from sky under gravity and being deployed at possible stable positions.}
		\label{fig:crystalgrowth}
	\end{center}
\end{figure}

Growth of Young diagrams is an important and interesting subject to study in mathematics. All the Young diagrams with $k$ boxes can be obtained by adding one box to all the diagrams with $k-1$ boxes in all possible allowed ways. Such a growth process is similar to the growth of a 2 dimensional crystal. See fig. \ref{fig:crystalgrowth}.
One can write a grand canonical partition function for this process \cite{Eynard:2008mt, Chattopadhyay:2019pkl}. Usually the growth depends on the probability of placing a crystal unit or box at different possible positions. A standard probability one assigns to such a growth process is called Plancherel probability. It is proportional to the square of the number of possibilities to obtain the shape of the crystal (Young diagram) with total $k$ number of units (boxes). It is easy to show that this number is equal to the dimension of the corresponding Young diagram. Vershik and Kerov \cite{VerKer77}, and independently, Logan and Shepp \cite{LogShe} studied such growth processes and showed that a growth under Plancherel measure terminates at a universal shape, when the diagrams are scaled appropriately as the number of boxes becomes large. The boundary of the terminal diagram becomes a smooth curve and follows the \emph{arc sin} law, known as the \emph{limit shape}. Kerov \cite{Kerov1} showed that the Young diagrams growing with the Plancherel transition probability follow a dynamical equation, called the \emph{automodel equation}. This equation is an inviscid Burger's equation. The limit shape is a unique solution of the automodel equation in far future with a Young diagram with no boxes as an initial condition in the far past.

The partition function of UMM admits a description in terms of representations of unitary group. In the large $N$ limit, there exists a correspondence between dominant eigenvalue distributions and Young diagrams associated with unitary representations. Such a relation allows a droplet description for different large $N$ phases of the matrix model under consideration. This was first observed in \cite{duttagopakumar} in the context of GWW model, and later in \cite{riemannzero,Chattopadhyay} for a generic class of UMM. From the shapes of these 2D droplets, one can find the dominant Young diagrams corresponding to the large $N$ phases of UMM. Thus, the evolution or growth of Young diagrams follows from the evolution of large $N$ phases of a UMM. It was shown in \cite{Chattopadhyay:2019pkl} that the Plancherel growth of Young diagrams is captured by GWW model. Evolution of no-gap phase of GWW model with respect to a parameter of the model exactly matches with the evolution of Young diagrams in the automodel class of Kerov. The limit shape was identified with GWW transition point. A Hilbert space description of Plancherel growth was also discussed in \cite{Chattopadhyay:2019pkl}. In this paper, we extend the analysis to \qq analog of Plancherel growth. 

There are different ways to consider \qdf of the Plancherel growth. The one that follows from the deformation of the Markov-Krein correspondence was considered by Strahov \cite{Strahov07}. In this construction, the probability associated to a Young diagram is proportional to the product of ordinary dimension and \qq dimension of the diagram. Strahov \cite{Strahov07} showed that the deformed inviscid Burger's equation (generalisation of automodel equation) has a fixed point and the fixed point is the \qd limit shape of Young diagrams. However, the explicit form of the limit shape was not given. There exists another class of \qd Plancherel probability which appears in different contexts in topological string theory. In this case, the $q$-probability of a Young diagram is proportional to the square of $q$-dimension of the diagram. One can write a partition function for such a growth process. The partition function is similar to that of topological string theory partition function on certain Calabi-Yau threefolds \cite{Caporaso:2006gk, Nekrasov:2003rj, Nekrasov:2002qd, Marino:2006hs} and Gromov-Witten invariants of $\mathbb{P}^1$ \cite{Eynard:2008mt, okounkov2006uses}. The $q$-Plancherel measure also appears in the partition function of $\mathcal{N}=2$ supersymmetric four dimensional gauge theory in $\Omega$-background, when expressed as a sum over partitions \cite{Nekrasov:2003rj}\footnote{Look at \cite{okounkov2006uses} and references therein for other applications of Plancherel measure and its $q$-deformation in physics.}. In this paper, we mainly focus on this particular growth process. The goal is to write a unitary matrix model which captures the \qd asymptotic growth of Young diagrams and set-up a mapping between the diagrams in the automodel class and the physical Hilbert space.\\
We summarise our main observations : 
\begin{itemize}
    
    \item We write down the partition function for a growth process equipped with \qd Plancherel measure considered in \cite{Caporaso:2006gk, Nekrasov:2003rj, Nekrasov:2002qd, Marino:2006hs,Eynard:2008mt}. One can exactly calculate the partition function using the normalisation properties of the measure. However, in order to get a handle on the asymptotic behaviour of the Young diagrams we introduce a ``fake" $N$ dependence in the problem, following \cite{Chattopadhyay:2019pkl}. We restrict the sum over representations to only those Young diagrams which have maximum $N$ number of rows. In case of ordinary Plancherel growth, the equivalence between the partition function and GWW model was exact and follows from the properties of characters of the symmetric group \cite{Chattopadhyay:2019pkl}. However, in the \qd version, the equivalence is not straight forward. We use the droplet picture of \cite{duttagopakumar, Chattopadhyay} to figure out the corresponding unitary matrix model. It turns out to be a $U(N)$ single plaquette model. We study the phase structure of the \qd partition function, in the large $N$ limit, which is analogous to that of the GWW model. The model exhibits a third order phase transition which is \qq analog of the GWW phase transition. The growth of Young diagrams under \qd Plancherel measure is captured by the gap-less phase of the UMM. The limiting or terminal diagram corresponds to the \qd GWW transition point. We also find the automodel class in this case and write the automodel equation satisfied by the members in this class. Following \cite{Chattopadhyay:2019pkl,Chattopadhyay:2020rle}, we construct the Hilbert space by quantising the phase space droplets and set up a correspondence between \qd automodel diagrams and coherent states in the Hilbert space. Finally, we elaborate the connection between moments of Young diagrams and the infinite number of commuting Hamiltonians obtained from the large $N$ droplets.
    
    \item We also discuss the growth process equipped with \qd Plancherel measure considered in \cite{Strahov07}.  In the large $N$ limit, the saddle point equation turns out to be a Riemann-Hilbert equation with a complicated kernel. We prescribe a method to solve such Riemann-Hilbert problem perturbatively in the deformation parameter. We obtain the dominant representations in the large $N$ limit corrected up to first order in the deformation parameter,
    and observe that both the growth processes are qualitatively same up to first order in the deformation parameter.
    
\end{itemize}

\section{$q$-deformed Plancherel growth}
\label{sec:qplancherel}

The Plancherel growth of Young diagrams has been a topic of intense study over many years. In order to explain the growth process, let $\lambda_k$ denote a Young diagram with $k$ boxes. Consider two Young diagrams $\lambda_k$ and $\lambda_{k+1}$ such that $\lambda_{k+1}$ is obtained from $\lambda_k$ by adding one box. One can associate a transition probability
\begin{equation}
\label{ptransition}
p_{\text{tran}}(\lambda_k;\lambda_{k+1})=\frac{1}{k+1}\frac{\dim \lambda_{k+1}}{\dim \lambda_k}
\end{equation}
if $\lambda_{k+1}$ is obtained from $\lambda_{k}$ by adding one box, and zero otherwise. Here, $\text{dim}\lambda_k$ denotes the dimension of the representation of symmetric group $S_k$ corresponding to Young diagram $\lambda_k$ and is given by
\begin{equation}
\label{eq:Frob-HL-formula}
\dim \lambda_k= \frac{k!} {\prod_{u\in \lambda_k}h(u)} =  \frac{k!}{\prod_i h_i!} \prod_{i<j}(h_i-h_j) .
\end{equation}
We denote the total number of boxes in a Young diagram by $l(\lambda)$ and the number of boxes in $i^{th}$ row by $\{\lambda_i\}_{i=1,..,l(\lambda)}$ such that $\lambda_1 \geq \lambda_2 \geq ... \geq \lambda_{l{(\lambda)}} \geq 0$. The hook number for a given $\lambda_i$ is given by
\ben
h_i=\lambda_i +l(\lambda)-i .
\een
They satisfy the constraint
\ben\label{eq:hiconst}
h_1>h_2> \cdots > h_{l(\lambda)} >0 .
\een
The expression \eqref{eq:Frob-HL-formula} is known as the \emph{hook length formula} \cite{fulton1991representation}. For a growth process governed by (\ref{ptransition}), one can calculate the probability to find a Young diagram $\lambda_k$ at the $k^{th}$ level, starting from a diagram with zero box. See \cite{Chattopadhyay:2019pkl} for details. It turns out to be the well known \emph{Plancherel probability} and is given by
\be
\label{pmeasure}
\mathcal{P}(\lambda_k)=\frac{(\text{dim}\lambda_k)^2}{k!}.
\ee

A natural $q$-deformation of the Plancherel growth process has been studied in \cite{Kerov1993AQO, Strahov07, 2010arXiv1001.2180F}. It follows from the representation theory of Iwahori-Hecke algebras. The transition probability of going from $\lambda_k$ to $\lambda_{k+1}$ is given by
\be
\label{q-transition}
p_{\text{tran}}^q(\lambda_k;\lambda_{k+1})=q^{b(\lambda_{k+1})-b(\lambda_k)}\frac{{\prod_{u \in \lambda_k}[h(u)]}}{{\prod_{u \in \lambda_{k+1}}[h(u)]}}
\ee
where $0<q<1$ is the deformation parameter. The constant $b$ is given by
\ben
b(\lambda) = \sum_{i=1}^{l(\lambda)}(i-1) \lambda_i
\een
and
\begin{equation}
\label{eq:boxidentity}
\frac{1}{\prod_{u \in \lambda_k}[h(u)]}=\frac{\prod_{i<j }[h_i-h_j]}{\prod_{i} [h_i]!}.
\end{equation}
The square bracket in the above expression represents the $q$-analogue of a positive integer defined as
\be\label{box-defn}
[x]=1-q^x 
\ee
and
\begin{equation}
[x]!=[x][x-1] \cdots [2][1].
\end{equation}
The probability measure associated to a Young diagram $\lambda_k$ growing according to (\ref{q-transition}) is given by
\be
\label{eq:Strahov-qPlancherel}
\mathcal{P}_q(\lambda_k)=(1-q)^k \dim \lambda_k \frac{q^{b(\lambda_k)}}{\prod_{u \in \lambda_k}[h(u)]} .
\ee
This is the $q$-deformed Plancherel measure considered in \cite{Strahov07, 2010arXiv1001.2180F}. Using $\lim_{q \to 1}\frac{[x]}{1-q}=x$, we see that
\begin{equation}
\label{eq:q-deformed-limit}
\lim_{q \to 1} \mathcal{P}_q(\lambda_k)=\mathcal{P}(\lambda_k) .
\end{equation}
Thus, in the limit $q\rightarrow 1$, the $q$-deformed measure reduces to the Plancherel measure.

\begin{figure}[h]
	\begin{center}
		\includegraphics[scale=0.16]{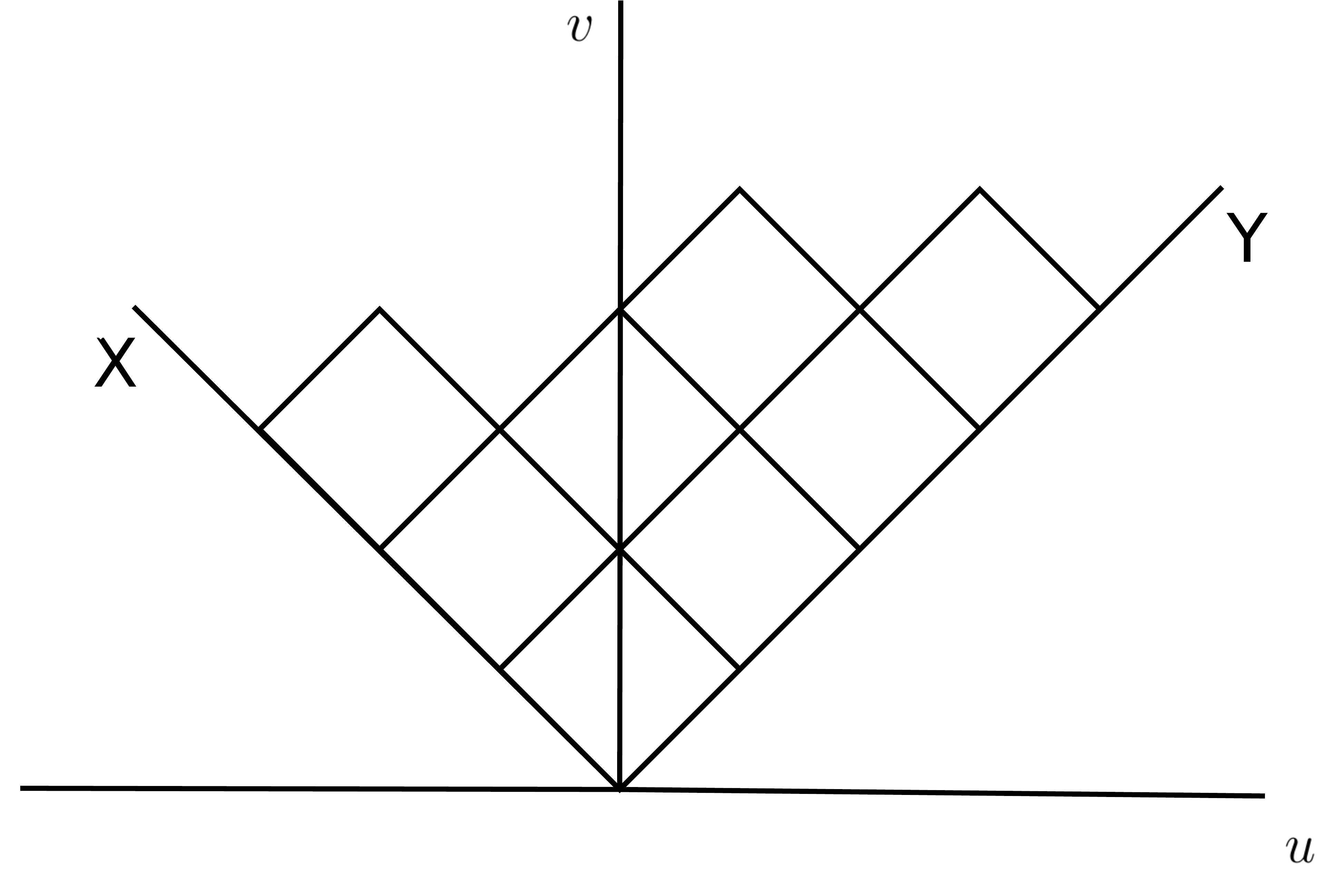}
		\caption{A typical Young diagrams written in the Russian convention.}
		\label{rotated_YD}
	\end{center}
\end{figure}

Strahov in \cite{Strahov07} gave a differential model for the $q$-deformed Plancherel growth process (\ref{q-transition}). He introduced a function $v_k(u)$ which describes the profile of a rotated  Young diagram as shown in fig.(\ref{rotated_YD}).
As the number of boxes becomes very large, Strahov defined a rescaled function
\be
\hat{v}_k(u)=\frac{v(u\sqrt{k})}{\sqrt{k}}
\ee
such that the area under the curve is finite. The boundary of Young diagram becomes smooth under scaling. He also introduced a \qd $R$ function associated with $\hat v_k$, given by
\be
R_{\hat{v}}(x;q)=\frac{1-q}{1-q^x}\text{exp}\left[-\ln q^{-1}\int \frac{d\sigma(u)}{1-q^{x-u}}\right]
\ee
where $\sigma(u)=\frac{1}{2}(\hat{v}(u)-|u|)$ is defined as the charge of a diagram \cite{Kerov1}. It was shown that the dynamical equation governing the growth of Young diagrams equipped with $q$-deformed Plancherel measure is given in terms of the $R$-function as
	\begin{equation}
	\label{q-burgers}
	\frac{\partial R_{\hat{v} (.,t;q)}(x;q)}{\partial x} + \frac{(1-q)}{\ln q^{-1}}R^{-1}_{\hat{v} (.,t;q)}(x;q)\frac{\partial R_{\hat{v} (.,t;q)}(x;q)}{\partial t} = 0.
	\end{equation}
This is a $q$-analogue of the automodel equation. It can be expressed in different equivalent forms in terms of various other quantities associated to Young diagrams \cite{Strahov07}. 

There exists another interesting class of \qq analog of Plancherel measure studied in \cite{Eynard:2008mt} for which
\begin{equation}\label{eq:qmeasure2}
 	\mathcal{P}_q(\lambda_k)= \left(\frac{\text{dim}_q\lambda_k}{k!}\right)^2 = k! \frac{(1-q)^{2k}\ q^{2b(\lambda_k)}}{\prod_{u \in \lambda_k} [h(u)]^2} =k!(1-q)^{2k}q^{2b(\lambda_k)}\frac{\prod_{1\leq i<j \leq N}[h_i-h_j]^2}{\prod_{i=1}^N ([h_i]!)^2}.  
\end{equation}
Such a measure plays an important role in the context of topological string theory partition function on certain Calabi-Yau threefolds and Gromov-Witten invariants of $\mathbb{P}^1$ \cite{Caporaso:2006gk, Nekrasov:2003rj, Nekrasov:2002qd, Marino:2006hs, okounkov2006uses}. Unlike (\ref{eq:Strahov-qPlancherel}), here the probability is proportional to the square of \qq dimension. This simplifies the large $N$ analysis to a great extent. The saddle point equation can be solved exactly in terms of the deformation parameter $q$. We observe that the behaviour of the automodel diagrams is qualitatively the same in both cases. 
Therefore, we mainly focus on the growth process governed by (\ref{eq:qmeasure2}) and write a $q$-analogue of the automodel equation in terms of \emph{Young diagram density}. For completeness, we also discuss the growth process under (\ref{eq:Strahov-qPlancherel}). However, in this case we are only able to provide the correction to automodel diagrams for the values of $q$ close to $1$.

\section{A unitary matrix model for $q$-deformed Plancherel growth}

The growth of Young diagrams can be compared with the growth of a two-dimensional crystal as shown in figure \ref{rotated_YD}. We consider every box as a unit falling from sky under gravity. There is a boundary given by the two lines passing through the origin with slope $\pm1$. Since the gravity is working downward it is easy to understand that the boxes will arrange themselves following the rules of a rotated Young diagrams. One can study a controlled growth process by assigning a probability to each diagram in this process. Following \cite{Eynard:2008mt, Chattopadhyay:2019pkl}, we can define a grand canonical ensemble of Young diagrams
\ben
\mathcal{E} = \bigcup_{k=0}^\infty \mathcal{Y}_k
\een
where $\mathcal{Y}_k$ is a set of Young diagrams with $k$ boxes. Next, assigning a probability for every diagram in the ensemble we write a grand canonical partition function for the $q$-deformed growth process as
\begin{equation}
\label{q-partition}
Z_q=\sum_{k=0}^{\infty}\sum_{\lambda_k} t^k \mathcal{P}_q(\lambda_k)\delta(k-|\lambda_k|)
\end{equation}
where $t>0$ is called \emph{fugacity}. In the above expression, $\mathcal{P}_q(\lambda_k)$ denotes the $q$-deformed Plancherel measure for a Young diagram $\lambda_k$ discussed in section \ref{sec:qplancherel}.

In order to study the above partition function in the large $k$ limit, we regularise the sum by imposing a cut-off on the Young diagrams in the summation over $\lambda_k$ in (\ref{q-partition}).  We introduce a large positive integer $N$ and constrain the Young diagrams in the ensemble to have no more than $N$ rows. This regularisation gives us a handle on the partition function to carry out a saddle point analysis of the problem. Later we show that the asymptotic behaviour of this regularised partition function can be captured by a unitary matrix model where $N$ plays the role of the rank of the unitary group.

The partition function \eqref{q-partition} for measure \eqref{eq:qmeasure2} is given by
\begin{equation}\label{eq:qdeformedgrowthpf2}
	\mathcal{Z}_q=\sum_{k=1}^{\infty}\sum_{\lambda_k}t^k \mathcal{P}_q(\lambda_k)=\sum_{k=1}^{\infty}\sum_{\lambda_k}t^k k! \frac{(1-q)^{2k}\ q^{2b(\lambda_k)}}{\prod_{u \in \lambda_k} [h(u)]^2}.
\end{equation}
It was observed in \cite{Chattopadhyay:2019pkl} that the ordinary Plancherel growth model is equivalent to $U(N)$ GWW model after restricting the Young diagrams to have maximum $N$ number of rows. The no-gap phase of GWW model describes the automodel class of Kerov. The limit shape corresponds to the GWW transition point. In this section, we first study the partition function (\ref{eq:qdeformedgrowthpf2}) in the large $N$ limit. The partition function admits two possible phases depending the value of the parameter  $t$ for a given $q$. The weak coupling phase of this theory captures the \qd growth of Young diagrams. Using the connection between UMM and 2D droplets we show that there exists an equivalent $U(N)$ UMM for the partition function (\ref{eq:qdeformedgrowthpf2}) whose no-gap phase captures the growth of Young diagrams under \qd Plancherel measure. Like GWW, this matrix model also admits a third order phase transition. This is a \qd analogue of GWW phase transition.

Using the definitions of \qq deformation (\ref{box-defn}), one can write the partition function as
\begin{equation}\label{eq:pfkind2}
	\begin{aligned}
	\mathcal{Z}_q & = \sum_{k=1}^{\infty}\sum_{\lambda_k}\exp \left[k \ln t + k \ln k-k+2k \ln (1-q)  + \sum_{i \neq j}\ln |q^{h_j}-q^{h_i}|\right. \\
	& \hspace{6cm} \left. - 2\sum_{i=1}^N \sum_{j=1}^{h_i}\ln (1-q^j)\right]
	\end{aligned}
\end{equation}
where we have dropped a constant term which makes no contribution to the saddle point equation.\\
In order to go to the continuum limit, we first re-define the parameter $q$,
\ben
q=e^{-g_s}
\een
and take the following double scaling limit
\begin{equation}\label{eq:doublescalling}
N \to \infty\ , \qquad g_s \to 0\ , \quad \mbox{such that}\ \lambda =N g_s\ \text{is finite.} 
\end{equation}
We also scale the variables and the summation appropriately with a factor of $N$ as follows
\begin{equation}
x=\frac{i}{N} , \quad  h(x) = \frac{h_i}{N}\quad \tand \quad \sum_{i=1}^{N} = N\int_0^1 dx.
\end{equation}
In the large $N$ limit, the total number of boxes scales as
\begin{equation}
k=\sum_{i=1}^N h_i -\frac{N(N-1)}{2} \xrightarrow[\text{limit}]{\text{large}\ N}N^2\int_0^1 dx\ h(x) -\frac{N^2}{2} \equiv N^2k'
\end{equation}
where
\begin{equation}
\int_0^1 dx\ h(x)=k'+\frac{1}{2}
\end{equation}
is an $O(1)$ quantity.\\
The partition function \eqref{eq:pfkind2} in the double-scaling limit can be written as
\begin{equation}
\cZ_q =\int [Dh]\, e^{S_{eff}},   
\end{equation}
where $S_{eff}$ is given by
\begin{equation}
	\begin{aligned}
	S_{eff}[h(x)]&=N^2 \left[k' \ln (N^2 t (1-q)^2 k')-k'+\int_0^1 dx \Xint-_0^1 dy \ln |e^{-\lambda h(x)}-e^{-\lambda h(y)}|\right.\\
	&\hspace*{7cm}\left. -2\int_0^1 dx \Xint-_0^{h(x)}dy \ln (1-e^{-\lambda y}) \right]
	\end{aligned}
\end{equation}
Varying the effective action with respect to $h(x)$, we get the saddle-point equation, which in terms of Young diagram density 
\ben
u(h)=-\frac{\partial x}{\partial h}
\een
is given by
\be
\label{eq:saddle_eynard}
\Xint-_{h_L}^{h_U} dh' u(h')\frac{e^{-\lambda h}}{e^{-\lambda h}-e^{-\lambda h'}}=-\frac{1}{\lambda} \log\left[\frac{1-e^{-\lambda h}}{\lambda\xi}\right]
\ee
with
\ben
\xi^2 = t \ k'.
\een
$u(h)$ has support between $h_L$ and $h_U$. From the monotonicity of $h_i$ (\ref{eq:hiconst}), it follows that the Young diagram density satisfies an upper bound
\ben
u(h) \leq 1.
\een
Therefore, one has to solve the above saddle-point equation in the presence of this constraint.\\
To solve the integral equation \eqref{eq:saddle_eynard}, we define new variables
\be\label{eq:hwrelation}
w = \frac{1-e^{-\lambda h}}{\lambda}
\ee
such that $u(w)= \frac{dh}{dw}u(h)$. 
In terms of these variables, the saddle point equation reduces to
\be
\label{eq:saddle_w}
\Xint-_{w_L}^{w_U} dw' \frac{u(w')}{w-w'} = \frac{1}{(1-\lambda w)} \log \frac{w}{\xi}
\ee
where $w_L=\frac{1-e^{-\lambda h_L}}{\lambda}$ and $w_U=\frac{1-e^{-\lambda h_U}}{\lambda}$ correspond to the end points of the support of $u(w)$.

Since the Young diagram density, $u(h)$ and hence $u(w)$ have an upper cap, we can classify solutions of (\ref{eq:saddle_w}) in two different classes. In the first class, the Young diagram density saturates the upper cap whereas in the second class, it does not. The solution class 1 captures the asymptotic growth of Young diagrams.

\subsection{Solution class 1: No-gap solution}
\label{sec:Plancherel-nogap}
\begin{figure}[h]
	\begin{center}
		\includegraphics[scale=0.6]{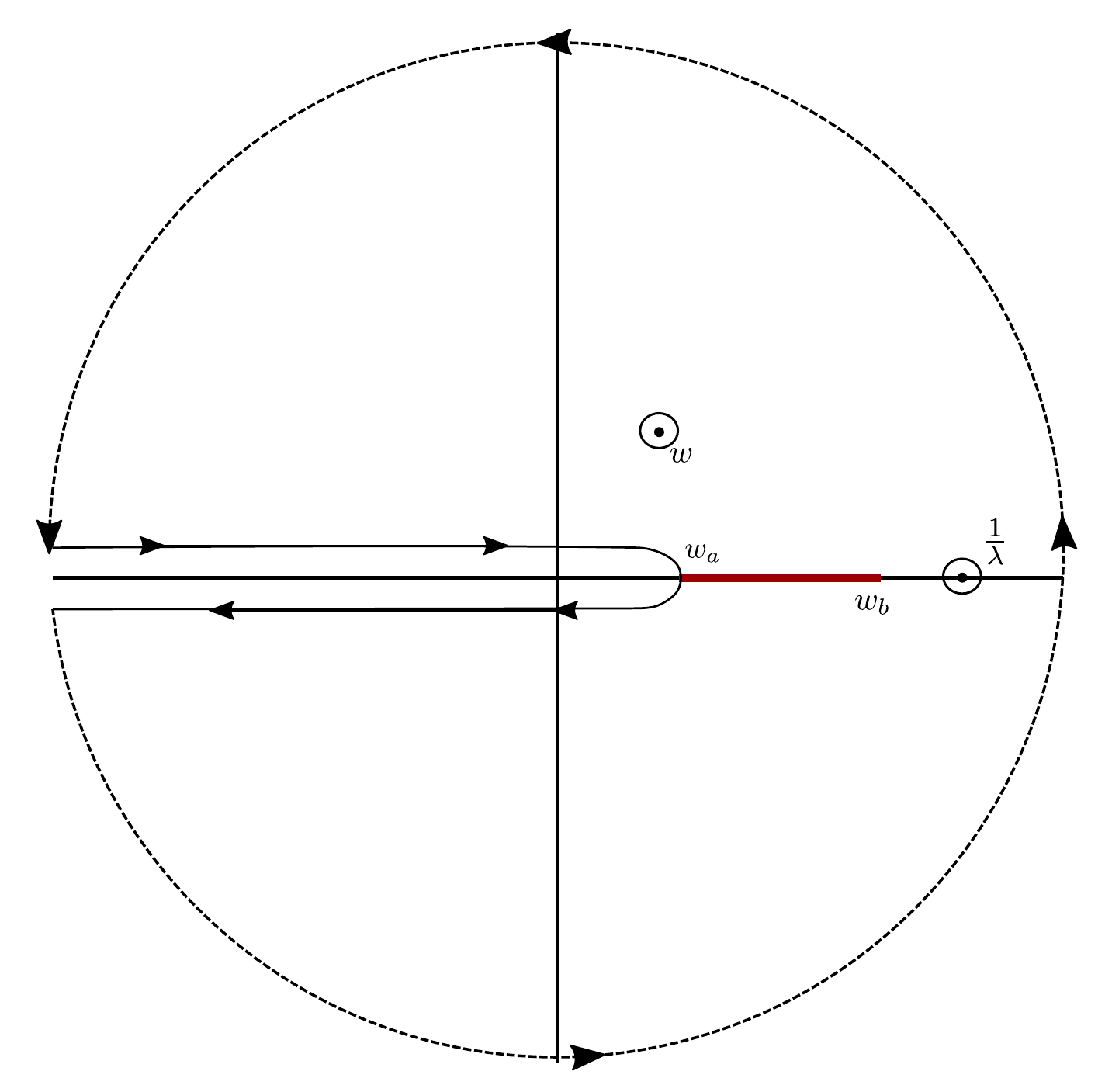}
		\caption{The deformed contour of integration for \eqref{eq:resolvent_w} has a logarithmic branch cut extending from $w_a$ to $-\infty$. There are first order poles at the points $s=w,\frac{1}{\lambda}$}
		\label{fig:w-contour}
	\end{center}
\end{figure}

Taking the ansatz $u(h)=1$ for $0\leq h\leq a$ and $u(h)=\tilde u(h)$ for $a\leq h\leq b$ in $h$-plane, we have
\be
u(w)=
\begin{cases}
\frac{1}{1-\lambda w} &\mbox{for}\  0\leq w\leq w_a\\
\tilde{u}(w) &\mbox{for}\ \ w_a\leq w\leq w_b 
\end{cases}
\ee
where $w_a = \frac{1-e^{-\lambda a}}{\lambda}$ and $w_b=\frac{1-e^{-\lambda b}}{\lambda}$. Putting this ansatz in the saddle point equation (\ref{eq:saddle_w}), we get
\be
\Xint-_{w_a}^{w_b} dw' \frac{\tilde{u}(w')}{w-w'} = \frac{1}{1-\lambda w}\log\left[\frac{w-w_a}{\xi(1-\lambda w_a)}\right].
\ee
We define a resolvent function
\ben\label{eq:def_resolvent}
H(w)=\int dw' \frac{u(w')}{w-w'},
\een
to solve the integral equation. $H(w)$ is analytic in the whole complex plane except at the support of $u(w)$, where it has a branch cut. The Young diagram density is given by the discontinuity of $H(w)$ about the branch cut. Following  the standard technique \cite{Migdal:1984gj}, we can write
\be
\label{eq:resolvent_w}
H(w) = \frac{1}{1-\lambda w}\log \frac{w(1-\lambda w_a)}{w-w_a} - \sqrt{(w-w_a)(w-w_b)}\oint_{\mathcal{C}}\frac{ds}{2\pi i}\frac{\log\left[\frac{s-w_a}{\xi(1-\lambda w_a)}\right]}{(s-w)(1-\lambda s)\sqrt{(s-w_a)(s-w_b)}}
\ee
where the contour $\mathcal{C}$ encloses the branch cut between $w_a$ and $w_b$. To evaluate the integral in (\ref{eq:resolvent_w}), we deform the contour as depicted in fig \ref{fig:w-contour}. 
The resolvent is finally given by
%
\be\label{eq:Hw}
\begin{aligned}
H(w) &= \frac{1}{1-\lambda w}\left[\log\frac{w}{\xi}+2\tanh^{-1}\sqrt{\frac{w-w_b}{w-w_a}}\right]\\ &\hspace*{3cm}+\frac{\lambda}{1-\lambda w} \sqrt{\frac{(w-w_a)(w-w_b)}{(1-\lambda w_a)(1-\lambda w_b)}}\left[\log\lambda\xi+ 2 \tanh^{-1}\sqrt{\frac{1-\lambda w_b}{1-\lambda w_a}}\right].
 \end{aligned}
\ee
From the asymptotic behaviour of the resolvent, one can find that the support of $\tilde{u}(w)$ is given by
\be
w_a = \frac{1-e^{-\lambda}(1+\lambda\xi)^2}{\lambda}, \quad w_b =\frac{1- e^{-\lambda}(1-\lambda\xi)^2}{\lambda} 
\ee
For $w$ inside the support, the discontinuity of $H(w)$ gives
\begin{equation}
\tilde{u}(w) = \frac{2}{\pi (1-\lambda w)} \tan^{-1}\sqrt{\frac{w_b-w}{w-w_a}}.
\end{equation}
In terms of our original variables $h(x)$, 
\begin{eqnarray}
    \tilde{u}(h)&=&(1-\lambda w)\,\tilde{u}(w)\nonumber\\
    &=&\frac{1}{\pi} \cos^{-1}\Big[\frac{1+\lambda^{2}\xi^{2}-e^{-\lambda(h-1)}}{2\lambda\xi}\Big]
    \label{q-automodel-D}
\end{eqnarray}
with support between $a$ and $b$ given by
\begin{equation}
    a=1-\frac{2}{\lambda}\log\left(1+\lambda\xi\right), \quad\text{and}\quad  b=1-\frac{2}{\lambda}\log\left(1-\lambda\xi\right).
\end{equation}
Since $a\geq 0$, the solution \eqref{q-automodel-D} is valid for
\begin{equation}
    \xi \leq \frac{1}{\lambda}(e^{\lambda/2}-1).
\end{equation}

\subsection{\qd automodel class}
\begin{figure}[h]
	\begin{center}
		\includegraphics[scale=0.4]{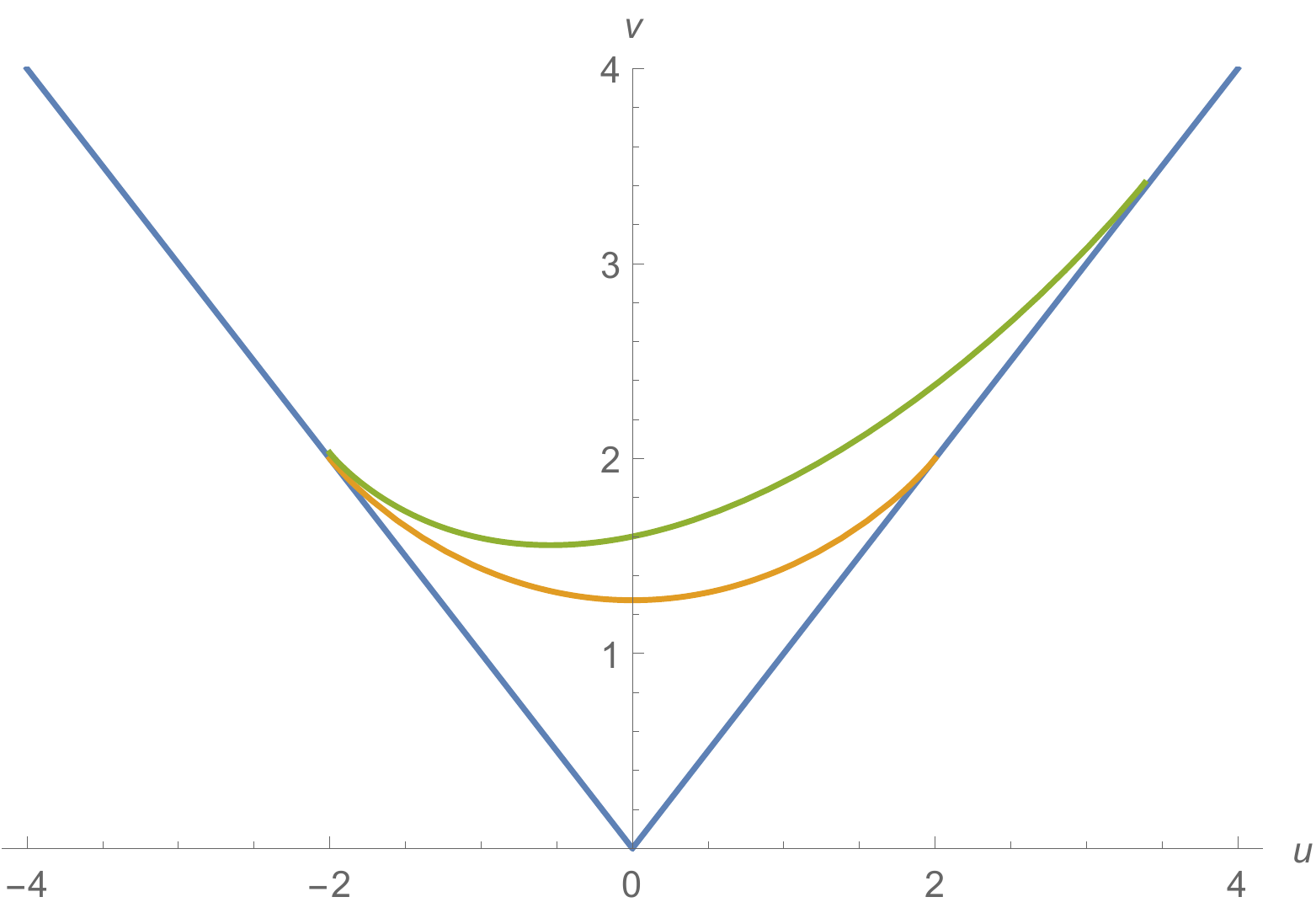}
		\caption{The limit shape for a $q$-deformed ($\lambda=0.8$) Plancherel growth process (green curve) and its $q\rightarrow 1$ ($\lambda\rightarrow 0$) limit (orange curve) for comparison.}
		\label{fig:q-limit-shape-2}
	\end{center}
\end{figure}

For $0<\xi\leq\frac{1}{\lambda}(e^{\lambda/2}-1)$, the density \eqref{q-automodel-D} corresponds to $q$-automodel diagrams which satisfy the following dynamical equation
\begin{equation}
\label{q-automodel-eqn}
    \partial_{\xi}\tilde{u}(h,\xi,\lambda)+\frac{\left[(1-\lambda^{2}\xi^{2})e^{\lambda(h-1)}-1\right]}{\lambda\xi}\partial_{h}\tilde{u}(h,\xi,\lambda)=0.
\end{equation}
This is a $q$-analogue of the automodel equation. One can translate this equation in terms of variables $v$ and $u$ through an appropriate change of variables defined in \cite{Chattopadhyay:2019pkl}. In $\lambda\rightarrow 0$ limit \eqref{q-automodel-eqn} reduces to the automodel equation
\begin{equation}
\label{eq:automodel-eqn}
    \partial_{\xi}\tilde{u}(h,\xi)+\frac{h-1}{\xi}\partial_{h}\tilde{u}(h,\xi)=0
\end{equation}
considered by Kerov \cite{Kerov1} and also derived by \cite{Chattopadhyay:2019pkl} in the language of Young diagram density. For the limiting value of $\xi$ i.e. $ \xi = \frac{1}{\lambda}(e^{\lambda/2}-1)$, the dominant Young diagram
\begin{equation}
\label{limit-shape}
 \tilde{u}(h)= \frac{1}{\pi}\cos^{-1}\left[ \frac{1-e^{-\lambda(h-1)}+(e^{\lambda/2}-1)^2}{2(e^{\lambda/2}-1)} \right]
\end{equation}
corresponds to the $q$-limit shape. It gives the limit shape for Plancherel growth process in the limit $\lambda\ra 0$. We plot the limit shape for both the Plancherel growth and it's $q$-deformation in fig \ref{fig:q-limit-shape-2}.
%

\subsection{The unitary matrix model from the droplet picture}

\label{sec:umm-droplet}
In order to give a Hilbert space description of the \qd growth process we need to find a UMM whose large $N$ phase can capture the growth of Young diagrams. For ordinary Plancherel growth it was shown in \cite{Chattopadhyay:2019pkl} that using the Frobenius formula one can write the partition function as GWW model. In case of \qd Plancherel growth (\ref{eq:qdeformedgrowthpf2}), it is tricky to use the Frobenius formula for \qd dimension. So, we take a different route. We use the droplet picture \cite{duttagopakumar,Chattopadhyay} to find the UMM that captures the asymptotic growth of Young diagrams under \qd Plancherel measure. Eynard \cite{Eynard:2008mt} also gave a matrix model for \qd Plancherel growth. The partition function was given by an integral over hermitian matrices with eigenvalues running over a circle of some specified radius. Similar connection was also pointed out in \cite{1998math.10105B}. Here, we explicitly find the unitary matrix model.

It was first observed in \cite{duttagopakumar} and subsequently by \cite{duttadutta,Chattopadhyay,Chattopadhyay:2019pkl} that the large $N$ phases of a generic UMM (known as single plaquette model), given by
\ben\label{eq:UMM}
\cZ = \int [DU] \exp\lB N\sum_{n=1}^\infty \frac{\beta_n}{n}\lb \Tr U^n+\Tr U^{\dagger n}\rb\rB,
\een
can be described in two different but equivalent ways - in terms of eigenvalue distribution and Young diagram distribution\footnote{It is well known that eigenvalues of unitary matrices behave like position of free fermions \cite{BIPZ} and the hook lengths of Young diagrams are like momenta of these fermions \cite{duttagopakumar,douglas2}.}. Different large $N$ phases can be characterised by both eigenvalue density and Young diagram distribution. Therefore, it is expected that there is a relation between these two descriptions. The relation is given by \cite{Chattopadhyay}
\ben\label{eq:hrhorel}
h^2-2S(\theta) h + S^2(\theta)-\pi^2 \rho^2(\theta) = 0
\een
where 
\ben\label{eq:Stheta}
S(\theta)=\frac12+\sum_n \beta_n \cos n\theta.
\een
The relation (\ref{eq:hrhorel}) allows us to provide a phase space picture for different large N phases of UMM
in terms of free fermi droplets (two dimensional distributions). These distributions/droplets are
similar to Thomas-Fermi distributions. Equation (\ref{eq:hrhorel}) has two possible solutions, $h_\pm(\theta)$
given by
\ben\label{eq:hpmtheta}
h_\pm(\theta) = S(\theta) \pm \pi \rho(\theta).
\een
Using this relation, one can define a distribution $\omega(h,\theta)$ in a two dimensional plane spanned by $(h,\theta)$
\be \label{eq:phasespacedistri}
\omega(h,\q) = \Theta\lb{(h-h_-(\q))(h_+(\q)-h)\over 2}\rb
\ee
such that the eigenvalue distribution can be obtained by integrating out $h$ for a given $\theta$
\be\label{eq:rhofromphasespace}
\r(\q)= \frac1{2\pi}\int dh \ \omega(h,\q) = {h_+(\q)-h_-(\q)\over 2\pi}.
\ee
The distribution also satisfies the normalisation condition
\ben\label{eq:omeganormalisation}
\frac1{2\pi}\int dh \ d\theta \ \omega(h,\theta) =1.
\een
The function $S(\q)$ can also be written in terms of phase space geometry as following
\be\label{eq:Sfromphasespace}
S(\q) = \frac1{2\pi \r}\int_0^\infty dh \ h \ \o(h,\q) =
{h_+(\q)+h_-(\q) \over 2}.
\ee
Integrating $\omega(h,\theta)$ over $\theta$ for a given $h$, one obtains a distribution of hook numbers $h$
\ben\label{eq:hdistribution}
u(h) = \frac1{2\pi}\int d\theta \ \omega(h,\theta)\quad \with \quad \int dh \ u(h) =1.
\een
If $\omega(h,\theta)$ has a single boundary for a given $h$ and $\theta >0$, then there is another identification between eigenvalue $\theta$ and the Young diagram density
\ben\label{eq:uhtheta}
\pi u(h) = \theta.
\een
In general, it is difficult to find the Young diagram distributions for different large $N$ phases of (\ref{eq:UMM}). However using the droplet picture one can obtain the dominant representations in the large $N$ limit for different phases of (\ref{eq:UMM}) \cite{Chattopadhyay}.

Contrary to \cite{Chattopadhyay}, here we know the large $N$ phases of (\ref{eq:qdeformedgrowthpf2}) in terms of dominant Young diagrams. Our goal is to find the corresponding eigenvalue distribution and hence the UMM. From the Young diagram distribution of automodel class (\ref{q-automodel-D}), we see that $u(h)$ is a smooth function of $h$. Inverting the relation, we find that $h$ and $\theta$ (using \ref{eq:uhtheta}) satisfy
\ben
h=1-\frac{1}{\lambda} \log\lB 1-2\lambda \xi \cos\theta+\lambda^2\xi^2 \rB.
\een
Comparing this with (\ref{eq:rhofromphasespace}) and (\ref{eq:Sfromphasespace}), we have
\ben
\label{eq:Stheta1}
\begin{split}
S(\theta) = \pi \rho(\theta) = \frac{h_+(\theta)}{2} & = \frac12-\frac{1}{2\lambda} \log\lB  1-2\lambda \xi \cos\theta+\lambda^2\xi^2 \rB \\
&= \frac12+ \sum_{n=1}^\infty \frac{\xi^n\lambda^{n-1}}{n} \cos n\theta
\end{split}
\een
and 
\ben
h_-(\theta) =0.
\een
Now, comparing (\ref{eq:Stheta1}) with (\ref{eq:Stheta}), we get
\ben\label{eq:betan}
\beta_n = \frac{\xi^n \lambda^{n-1}}{n}.
\een

In order to complete the story, we note that the regularised partition function (\ref{eq:qdeformedgrowthpf2}) admits a phase transition in the large $N$ limit at \ben
\xi = \frac{1}{\lambda}(e^{\lambda/2}-1).
\een
For $\xi > \frac{1}{\lambda}(e^{\lambda/2}-1)$, the Young diagram density does not saturate the upper bound and is given by (see \ref{app:qGWWPT} for derivation)
\be
\label{eq:uh_gapped_phase}
u(h) =\frac{1}{\pi } \cos^{-1}\left[\frac{ 1+e^{\lambda} e^{-2 \lambda h} -2 e^{\lambda/2}e^{-\lambda h}(1+\lambda\xi)+\lambda\xi(2e^{-\lambda h} +\lambda\xi)}{2\lambda (1-e^{-\lambda h}) \xi}\right].
\ee
We need to confirm that the dictionary (\ref{eq:rhofromphasespace}) and (\ref{eq:Sfromphasespace}) renders the same UMM in this phase of the theory. Inverting \eqref{eq:uh_gapped_phase}, we see that there exist two possible solutions for $h$ for a given $\pi u(h)=\theta$. It is easy to check using (\ref{eq:Sfromphasespace}) that we obtain the same $S(\theta)$ (see \ref{app:qGWWPT} for details). Thus, we conclude that the asymptotic (large $N$) structure of the \qd partition function (\ref{eq:qdeformedgrowthpf2}) is captured by a UMM (\ref{eq:UMM}) with $\beta_n$ given by (\ref{eq:betan}). The \qd growth of Young diagrams is captured by the no-gap phase of this UMM.

\section{Asymptotic analysis of $q$-deformed Plancherel growth of second kind}
\label{subsec:qdkind1}

In this section, we discuss the growth process controlled by the \qd probability measure considered in \cite{Strahov07}, given by \eqref{eq:Strahov-qPlancherel}. The partition function \eqref{q-partition} can be written as
\begin{equation}\label{eq:PF_Strahov}
Z_q=\sum_{k=1}^{\infty}\sum_{\lambda_k} \exp[S_{eff}(h_i)]e^{-\frac{\ln q}{6}N(N-1)(N-2)}
\end{equation}
where the effective action is given by (see \ref{app:calculationofZ})
\begin{equation}
\label{eff-S}
\begin{aligned}
S_{eff}(h_i) = k\ln t(1-q) +k\ln k -k & + \sum_{ i\neq j} \frac{1}{2} \ln |h_i-h_j|-\sum_{i} (h_i \ln h_i -h_i) \\
& + \sum_{ i\neq j}  \frac{1}{2}\ln |q^{h_j}-q^{h_i}| -\sum_{i=1}^{N} \sum_{j=1}^{h_i} \ln (1-q^j).
\end{aligned}
\end{equation}
In the continuum limit, the effective action takes the form
\begin{equation}
\begin{aligned}
S_{eff}[h(x)]&=N^2\left[k'\ln t(1-q)N+k' \ln k' -k'-\int_0^1 dx\ (h(x)\ln h(x)-h(x))\right.\\
&\hspace*{-1.7cm}\left.+\frac{1}{2}\int_0^1 dx \Xint-_{0}^{1} dy \left(\ln |q^{Nh(x)}-q^{Nh(y)}|
+\ln|h(x)-h(y)|\right)-\int_0^1 dx \int_0^{h(x)}dy\ \ln(1-q^{Ny})\right].
\end{aligned}
\end{equation}
In the double scaling limit (\ref{eq:doublescalling}), the dominant contribution to the partition function comes from the configurations satisfying the saddle point equation
\begin{equation}
\label{fulleom}
\Xint-_{h_{L}}^{h_{U}} dh' u(h')\left[\coth \frac{\lambda(h-h')}{2}+\frac{2}{\lambda(h-h')}\right]=1+\frac{2}{\lambda}\ln \frac{h(1-e^{-\lambda h})}{t\lambda k'}
\end{equation}
where $[h_L,h_U]$ denotes the support of $u(h)$. Our goal is to find a solution of (\ref{fulleom}) for $\lambda\neq 0$.\\
In order to solve the saddle point equation (\ref{fulleom}), we take the following ansatz for $u(h)$
\begin{equation}
\label{eq:u_ansatz}
u(h)=
\begin{cases}
1 &\mbox{for}\ 0\leq h\leq a \\
\tilde{u}(h) &\mbox{for}\  a\leq h\leq b.
\end{cases}
\end{equation}
The equation of motion \eqref{fulleom} takes the form
\begin{equation}
\label{eom2}
\Xint-_{a}^{b} dh'\ \tilde{u}(h')\left[\coth \frac{\lambda(h-h')}{2}+\frac{2}{\lambda(h-h')}\right]=1-a+\frac{2}{\lambda}\ln \left[\frac{(h-a)(1-e^{-\lambda (h-a)})}{\lambda \xi^2}\right].
\end{equation}
The kernel in the above integral equation is nontrivial. It is difficult to do a change of variable which brings the kernel to a standard form. However, the $\coth$ part of the kernel has the same singular structure as that of the second part. We therefore expand the $\coth$ part in powers of $\lambda$ and separate out the singular part. The derivation is given in \ref{app:strahov}. To find $\tilde{u}(h)$, we again define a resolvent function as in \eqref{eq:def_resolvent}. We provide the explicit calculations for the resolvent in \ref{app:strahov}. Here, we present the final answer.
\begin{eqnarray} 
\label{resolvent-exact}
\begin{split}
 H(h) = \ln \frac{h}{\xi} & + \ln \frac{\sqrt{b}-\sqrt{a}}{\sqrt{b}+\sqrt{a}}+\ln \frac{h + \sqrt{a b} - \sqrt{(h-a)(h-b)}}{h-\sqrt{a b} + \sqrt{(h-a)(h-b)}} + \frac{\lambda}{4}(1-a)+\sum_{k=1}^{\infty} A_{k}^{(1)}\sqrt{\frac{(h-a)}{(h-b)}} \\ & + \sum_{k=1}^{\infty}A_{k}^{(2)}\frac{_2F_1(1,\frac{1}{2}+k,2+k,\frac{a-b}{a-h})}{\sqrt{(h-a)(h-b)}}+\sum_{k=1}^{\infty}\sum_{n=0}^{2k-1}A_{k,n}^{(1)}\sqrt{\frac{(h-a)}{(h-b)}} \\ & + \sum_{k=1}^{\infty}\sum_{n=0}^{2k-1}A_{k,n}^{(2)}\frac{_2F_1(1,\frac{1}{2}+n,2+n,\frac{a-b}{a-h})}{\sqrt{(h-a)(h-b)}}       
\end{split}
\end{eqnarray}
where $A_k$'s are given by equation (\ref{eq:Aks}). Different moments, $\mathfrak{h}_n$ (defined in \eqref{shifted-moment}) appear on the right hand side of this expression, which are a priori not known. Hence, the saddle point equation has to be solved recursively. However, if we calculate the resolvent order by order in $\lambda$, the moments start appearing only at order $\lambda^2$. The result up to order $\lambda$ is given by
\begin{equation}
\label{resolvent-approx}
\begin{aligned}
 H(h)&=\ln \frac{h}{\xi}+\ln \frac{\sqrt{b}-\sqrt{a}}{\sqrt{b}+\sqrt{a}}+\ln \frac{h+\sqrt{ab}-\sqrt{(h-a)(h-b)}}{h-\sqrt{ab}+\sqrt{(h-a)(h-b)}}+\frac{\lambda}{4}(1-h)\\&\hspace{1cm} + \frac{\lambda\sqrt{(h-a)(h-b)}}{4}.
\end{aligned}
\end{equation}
The support of $\tilde{u}(h)$ can be found from the asymptotic behaviour of the resolvent. Up to first order in $\lambda$, we get
\begin{eqnarray}
a = 1-2\xi +\lambda\frac{\xi^{2}}{2},\quad
b = 1+2\xi +\lambda\frac{\xi^{2}}{2}.
\end{eqnarray}
and the Young diagram density is given by
\begin{equation}\label{eq:tildeuhstrahov}
\tilde{u}(h)=\frac{1}{\pi}\cos^{-1}\Big(\frac{2h-a-b}{b-a}\Big)-\frac{\lambda\sqrt{(h-a)(b-h)}}{4\pi}.
\end{equation}
Since $a\geq 0$, the above solution is valid for 
\be\label{eq:xilimitstrahov}
0<\xi\leq \frac12 +\frac{\lambda}{16}.
\ee
$\tilde u(h)$ given by (\ref{eq:tildeuhstrahov}) describes the automodel class of \cite{Kerov1} corrected up to first order in $\lambda$. It is easy to check from equations (\ref{q-automodel-D}) and (\ref{eq:tildeuhstrahov}) that the order $\lambda$ correction to automodel density in both the cases is same apart from a factor of half. The reason behind this is the appearance of $\text{dim}\lambda_k$ in (\ref{eq:Strahov-qPlancherel}). 

\begin{figure}[H]
	\begin{center}
		\includegraphics[scale=0.4]{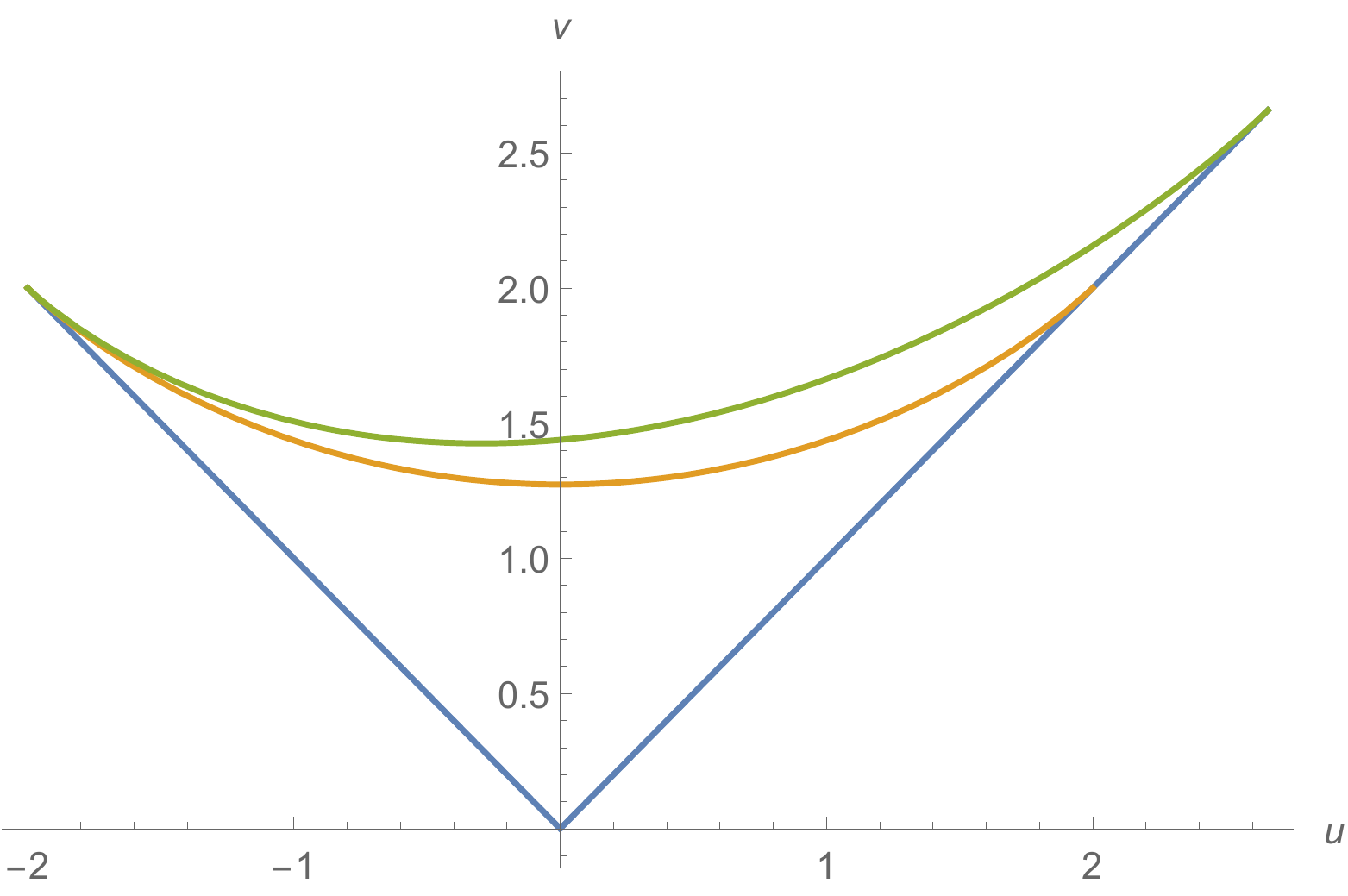}
		\caption{The limit shape for a $q$-deformed ($\lambda=0.97$) Plancherel growth process (green curve) and its $q\rightarrow 1$ ($\lambda\rightarrow 0$) limit (orange curve) for comparison.}
		\label{fig:strahov-ls}
	\end{center}
\end{figure}

The limit shape corresponds to the upper limit of $\xi$ given in (\ref{eq:xilimitstrahov}). We plot the limit shape in a rotated coordinate system (see \cite{Chattopadhyay:2019pkl} for coordinate transformation) in fig \ref{fig:strahov-ls}. It is clear that the symmetric property of the distribution \cite{Chattopadhyay:2019pkl} is lost under \qq deformation. 
Thus, we see that both the measures (\ref{eq:Strahov-qPlancherel} and \ref{eq:qmeasure2}) qualitatively give the similar asymptotic behaviour up to order $\lambda$. However, we emphasise that (\ref{q-automodel-D}) is exact in $\lambda$.

\section{The Hilbert space}
\label{sec:hilbertspace}

Following \cite{ Chattopadhyay:2020rle,Chattopadhyay:2019pkl} we give a Hilbert space description of the \qd growth process. Using the connection between unitary matrix model and free Fermi droplet description, we map the Young diagrams in automodel class to different shapes of two dimensional phase space droplets. Quantising these droplets, we further set up a correspondence between automodel diagrams and coherent states in the Hilbert space. Thus, growth of Young diagrams are mapped to evolution of coherent states in the Hilbert space. The basic construction of the Hilbert space is similar to the one considered in \cite{Chattopadhyay:2019pkl}. We briefly review the construction in \ref{app:hi}.

The Hilbert space $\cH$ is a direct product of $\cH_+$ and $\cH_-$ where $\cH_\pm$ are the Hilbert spaces associated with the $+$ and $-$ sectors defined in section \ref{sec:umm-droplet} namely the upper and the lower fermi surfaces. These two sectors are isomorphic. A generic state in $\cH_+$ is given by
\ben
\ket{\vec k} = \prod_{n\geq 1} (a^\dagger_n)^{k_n}\ket{s}
\een
where $a^\dagger_n$s are creation operators in $\cH_+$ and $\ket{s}$ is the ground state annihilated by $a_n$s. $\ket{\vec k}$ satisfies the completeness relation (\ref{eq:completeness}). Therefore, $(1/\sqrt{z_{\vec k}})\ket{\vec k}$ is the normalised excited state. One can also define coherent states in the Hilbert space $\cH_+$,
\ben
\ket{\tau_+} = \exp\lb \sum_{n=1}^\infty\frac{\tau_n^+ a_n^\dagger}{n\hbar}\rb \ket{s}. 
\een
%
We set up a map between a state $\ket{\psi} \in \cH_+$ and shape of the upper Fermi surface
\ben\label{eq:maping}
\ket{\psi} \ra \{\bra{\psi}\bar{h}_+(\theta)\ket{\psi}\}.
\een
%
Such a mapping has been discussed in \cite{Chattopadhyay:2020rle}. Expectation value of $\bar h_+$ in ground state is given by $\frac12 + s \hbar$ and has zero dispersion. Thus, the ground state $\ket s$ corresponds to an overall shift of $\bar h_+$ by an amount $s/N$ over the classical value. Generic \emph{normalised} excited states correspond to ${\cal O}(\hbar)$ ripples on $\bar h_+$. In this case, the expectation value of $\bar{h}_+$ picks up a non-zero dispersion at ${\cal O}(\hbar)$. ${\cal O}(1)$ deformations of the upper Fermi surface $\bar{h}_+$ are mapped to coherent states $\ket{\tau_+}$ in $\cH_+$ as is evident from (\ref{eq:coherentprofile}). An important thing to note here is that the classical deformations do not disturb the quadratic profile of the droplets \emph{i.e} for a given $\theta$, there exists unique values of $h_\pm(\theta)$.

We first consider the state corresponding to automodel diagrams for ordinary Plancherel growth \cite{Chattopadhyay:2019pkl}. The automodel diagrams corresponds to
\ben
\bar h_\pm(\theta) = \pm \lb \frac12 +\xi \cos\theta\rb. 
\een
The set of coherent states with $\tau_1=\xi/2$ , $\tau_n = 0$ $\forall n > 1$ are mapped to the automodel diagrams. In particular, the state corresponding to the limit shape is given by
\ben
|\tau* \rangle = \text{exp}\left(\frac{a_{1}^{\dagger}}{2\hbar}\right) |0\rangle. \quad
\een

Our final goal is to find the coherent state associated with automodel diagrams for \qd Plancherel growth. For the \qd automodel class, the variable $\bar h_+(\theta)$ is given by
\begin{equation}
    \bar{h}(t,\theta)=\frac{1}{2}-\frac{1}{2\lambda}\log(1-2\lambda \xi \cos \theta +\lambda^2 \xi^2)=\frac12 + \sum_{n=1}^{\infty}\frac{\xi^n \lambda^{n-1}}{n}\cos n\theta. 
\end{equation}
Therefore the coherent state associated to q-automodel class is given by
\ben
\ket{\tau^+_q} = \exp\lb {\sum_{n=1}^\infty \frac{\t_{n}^+ a_n^\dagger}{n\hbar}}\rb \ket{s},\quad \where \quad \tau^+_n = \frac{\xi^n \lambda^{n-1}}{n}.
\een

\subsection{Moments of Young diagrams and conserved charges}
Following \cite{Jevicki:1993qn} one can define an infinite sequence of conserved commuting quantities 
\ben
H_n = \frac{1}{2\pi} \int dh\ d\theta\ \omega(h,\theta) \ h^n.
\een
Using the Poisson bracket (\ref{eq:poisson}) one can show that 
\ben
\{H_m,H_n\} = 0.
\een
Also, using the equations of motion (\ref{eq:boundeq}) one can show that %
\ben
\frac{d H_n}{dt} =0. 
\een
Existence of infinite number of conserved charges makes the system integrable. This is a generic feature of one matrix model and indicates that the theory is exactly solvable.

These conserved charges are related to moments of Young diagrams defined in \cite{Kerov1}. The matrix model techniques can be used to calculate the moments of automodel diagrams. Using the definitions (\ref{eq:hdistribution}), one can show that
\ben
H_n = \int dh \ u(h)\ h^n.
\een
One can define a generating function for these moments
\ben
H(h) = \int\ dh' \frac{u(h')}{h-h'} = \sum_{n=0}^\infty \frac{H_n}{h^{n+1}}.
\een
Using the expression for the resolvent (\ref{eq:Hw}) and the relation between $h$ and $w$ (\ref{eq:hwrelation}), one can show that
\ben
H_n = \sum_{i_1,\cdots, i_n \geq 1} \frac{\lambda^{i_1+\cdots +\i_n-n}}{i_1\cdots i_n} p_{i_1+\cdots+i_n}
\een
where $p_n$s are the moments associated with $H(w)$,
\be
H(w) = \sum_{n=0}^\infty \frac{p_n}{w^{n+1}}.
\ee
Since we know the exact expression for $H(w)$, one can easily find the $p_n$s from it's asymptotic expansion. In the limit $\lambda\ra 0$, we get 
\ben
H_n=p_n.
\een
Thus, we have an explicit formula to compute the moments of the asymptotic automodel diagrams.

\section{Conclusion}

In this paper, we discuss about \qd Plancherel growth of Young diagrams. We construct a unitary matrix model that captures the growth of Young diagrams with the coupling constant playing the role of time \cite{Strahov07}. The matrix model is a \qq analog of GWW model and the growth of Young diagrams equipped with a \qd Plancherel measure is captured by the no-gap phase of this model. The \qd limit shape is the \qd GWW transition point. We use the connection between unitary matrix model and 2D droplet picture of \cite{duttagopakumar,Chattopadhyay} to construct the unitary matrix model. Construction of this matrix model allows us to map the automodel diagrams to different shapes of the 2D droplets. Upon quantising the droplets, one can construct a Hilbert space and the automodel diagrams can be mapped to coherent states in that Hilbert space. We study two different $q$-deformations of the Plancherel measure. One follows from the Iwahori-Hecke algebras, considered in \cite{Strahov07,2010arXiv1001.2180F}. In this case, the measure is proportional to a product of \qd dimension and ordinary dimension of the representation. The other kind of \qd measure is proportional to the square of \qq dimension. Such measure appears in different contexts in topological string theories and gauge theories. We see that in the large $N$ limit, the later can be solved exactly in the deformation parameter, while we do a perturbative analysis of the former. However, the automodel diagrams in either case have the same qualitative behaviour for small deformation.

We compute the moments of different automodel diagrams explicitly and show that these moments are related to the infinite number of commuting conserved charges associated with the one matrix model. The moments of Young diagrams can be obtained from the large $h$ expansion of resolvent. Since the saddle point technique allows us to compute the resolvent exactly in $\lambda$, in the large $N$ limit, we have been able to compute the moments of automodel diagrams explicitly.

\vspace{1cm}

\noindent{\bf Acknowledgments:} The work of SD is supported by the MATRICS grant no.          \\    
MTR/2019/000390 from the Department of Science and Technology,
Government of India. DM, N and SP acknowledge many illuminating discussions with Arghya Chattopadhyay. Neetu would like to thank Arindam Bhattacharya for many fruitful discussions over the course of this work. We also thank all the medical and non-medical workers who are working tirelessly in these troubled times. Finally, we are grateful to the people of India for their unconditional support towards researches in basic sciences.

\appendix

\section{\qq analog of GWW phase transition}\label{app:qGWWPT}
The no-gap phase of the \qd Plancherel growth process captured by the partition function \eqref{eq:qdeformedgrowthpf2} has been discussed in section \ref{sec:Plancherel-nogap}. 
There exist another class of solution for the saddle point equation \eqref{eq:saddle_eynard}, namely the one-gap solution for which 
\be
0\leq u(h)< 1 \quad\text{for}\quad h\in [a,b]\ .
\ee
To solve the saddle point equation for this case, we define new variables
\be
z=\frac{1-e^{-\lambda h}}{\lambda}
\ee
in terms of which the equation becomes
\be
\label{saddle_z_nogap}
\Xint-_{z_p}^{z_q} dz'\,\frac{u(z')}{z-z'}=\frac{1}{(1-\lambda z)}\log\left(\frac{z}{\xi}\right)
\ee
where $z_a$ and $z_b$ correspond to points $a$ and $b$ in the $h$ plane.
The resolvent for the above integral equation is given by
\be
H(z)=-\sqrt{(z-z_a)(z-z_b)}\oint_{\mathcal{C}}\frac{ds}{2\pi i}\frac{\log(s/\xi)}{(1-\lambda s)(s-z)\sqrt{(s-z_a)(s-z_b)}}
\ee
where similar to the previous cases, the contour $\mathcal{C}$ encloses the branch cut between $z_a$ and $z_b$. We deform the contour as shown in fig \ref{fig:w-contour}. Taking contributions from all the poles and the logarithmic branch cut, the resolvent is finally given by
\be
\begin{aligned}
H(z)=&-\frac{2}{(1-\lambda z)} \log\left[\frac{\left(\sqrt{z_a(z_b-z)}-i\sqrt{z_b(z-z_a)}\right)\sqrt{\xi}}{z(\sqrt{z_b-z}-i\sqrt{z-z_a})}\right]\\
&-\frac{2\lambda}{(1-\lambda z)} \frac{\sqrt{(z-z_a)(z-z_b)}}{\sqrt{(1-\lambda z_a)(1-\lambda z_b)}}\log\left[\frac{\sqrt{z_a(1-\lambda z_b)}-\sqrt{z_b(1-\lambda z_a)}}{\left(\sqrt{1-\lambda z_b}-\sqrt{1-\lambda z_a}\right)\sqrt{\xi}}\right]
\end{aligned}
\ee
The discontinuity of the resolvent gives
\be
u(z)=\frac{1}{\pi (1-\lambda z)} \cos^{-1}\left[\frac{4\sqrt{z_a z_b}(z - z_a)(z_b - z)+(z_a + z_b -2z)(2 z_a z_b - z z_a- z z_b)}{(z_b - z_a)^2 z}\right]
\ee
The support of $u(z)$ is given by
\be
\begin{aligned}
\sqrt{z_a}=e^{-\lambda/2}\sqrt{\xi} - \frac{e^{-\lambda/2}\sqrt{-e^{-\lambda/2}+e^{\lambda}+\lambda\xi-e^{-\lambda/2}\lambda\xi}}{\sqrt{\lambda}} \\
\sqrt{z_b}=e^{-\lambda/2}\sqrt{\xi} + \frac{e^{-\lambda/2}\sqrt{-e^{-\lambda/2}+e^{\lambda}+\lambda\xi-e^{-\lambda/2}\lambda\xi}}{\sqrt{\lambda}}
\end{aligned}
\ee
In terms of our original variable $h$, we have
\be
u(h) =\frac{1}{\pi } \cos^{-1}\left[\frac{ 1+e^{\lambda} e^{-2 \lambda h} -2 e^{\lambda/2}e^{-\lambda h}(1+\lambda\xi)+\lambda\xi(2e^{-\lambda h} +\lambda\xi)}{2\lambda (1-e^{-\lambda h}) \xi}\right].
\ee

Inverting this relation with the help of the identification (\ref{eq:uhtheta}), we find
\ben
\begin{split}
h_+(\theta) &= \frac{\log \left(\frac{-\lambda  \xi  \cos (\theta )+\sqrt{\lambda  \xi  (\cos (\theta
   )+1) \left(\lambda  \xi  (\cos (\theta )+1)-2 e^{\lambda /2} (\lambda  \xi +1)+2
   e^{\lambda }\right)}-\lambda  \xi +e^{\lambda /2} (\lambda  \xi +1)}{-2 \lambda  \xi 
   \cos (\theta )+\lambda ^2 \xi ^2+1}\right)}{\lambda }\\
h_-(\theta) &= \frac{\log \left(\frac{\lambda  \xi  \cos (\theta )+\sqrt{\lambda  \xi  (\cos (\theta
   )+1) \left(\lambda  \xi  (\cos (\theta )+1)-2 e^{\lambda /2} (\lambda  \xi +1)+2
   e^{\lambda }\right)}+\lambda  \xi -e^{\lambda /2} (\lambda  \xi +1)}{2 \lambda  \xi 
   \cos (\theta )-\lambda ^2 \xi ^2-1}\right)}{\lambda }.
\end{split}
\een
Now using the mapping (\ref{eq:Sfromphasespace}), we find $S(\theta)$ which matches with (\ref{eq:Stheta1}).

\section{Explicit computation of the partition function}
\label{app:calculationofZ}
Here, we provide some details of the computation for the partition function \eqref{eq:PF_Strahov} that follows if one takes the \qd Plancherel measure following Strahov \cite{Strahov07}. The partition function can be written as
\begin{equation}
\begin{aligned}
Z_q
&=\sum_{k=1}^{\infty}\sum_{\{h_i\}} \exp\left[k\ln t+k \ln (1-q)+b(\lambda_k) \ln q +\ln k! \right] \\
&\hspace*{2cm} \exp \left[\ln \prod_{1 \leq i<j\leq N}(h_i-h_j)-\ln \prod_{i=1}^Nh_i! +\ln \prod_{1 \leq i<j\leq N}[h_i-h_j]-\ln \prod_{i=1}^N[h_i]! \right].
\end{aligned}
\end{equation}
The total number of boxes in a Young diagram can be expressed as 
\ben
k=\sum_{i=1}^N \lambda_i=\sum_{i=1}^N (h_i-N+i)=\sum_{i=1}^N h_i -\frac{N(N-1)}{2}.
\een
We will now write down the above in the large $N$ limit. Large $N$ naturally implies large $k$ which is an $O(N^2)$ number. We can also write
\begin{equation}
b(\lambda_k)=\sum_{i=1}^N (i-1)(h_i-N+i)=\sum_{i=1}^N (i-1)h_i-\frac{N(N-1)(N-2)}{6}\ .
\end{equation}
Using Stirling's approximation, the partition function in the large $N$ limit becomes
\begin{equation}
\begin{aligned}
Z_q &\approx \sum_{k=1}^{\infty}\sum_{\lambda_k} \exp \left[k\ln t(1-q) + \sum_{i=1}^N (i-1)h_i \ln q +k\ln k -k +\frac{1}{2} \sum_{1 \leq i\neq j\leq N} \ln |h_i-h_j|\right.\\
&\left. -\sum_{i=1}^N(h_i \ln h_i -h_i)+\sum_{1 \leq i<j\leq N} \ln [h_i-h_j] -\sum_{i=1}^N \ln [h_i]!\right]\exp\left[-\frac{\ln q}{6}N(N-1)(N-2)\right]\ .
\end{aligned}
\end{equation}
We can simplify the last two terms appearing in the exponent of above expression as follows:
\begin{equation}
\label{T3exprn}
\begin{aligned}
\sum_{1 \leq i<j\leq N} \ln [h_i-h_j]&=\sum_{1 \leq i<j\leq N} \ln (1-q^{h_i-h_j})=\sum_{1 \leq i<j\leq N} \left(\ln(q^{h_j}-q^{h_i}) -h_j \ln q\right)\\
&= \frac{1}{2} \sum_{1 \leq i\neq j\leq N}\ln |q^{h_j}-q^{h_i}|-\sum_{i=1}^N (i-1)h_i \ln q
\end{aligned}
\end{equation}
and
\begin{equation}
\label{T4exprn}
\begin{aligned}
\sum_{i=1}^N \ln [h_i]!&=\sum_{i=1}^{N}\sum_{j=1}^{h_i} \ln [j]=\sum_{i=1}^{N}\sum_{j=1}^{h_i} \ln (1-q^j)\ .
\end{aligned}
\end{equation}
Using \eqref{T3exprn} and \eqref{T4exprn} in the expression of the partition function, we end up with
\begin{equation}
\begin{aligned}
Z_q &\approx \sum_{k=1}^{\infty}\sum_{\lambda_k} \exp \left[k\ln t(1-q) +k\ln k -k +\sum_{1 \leq i\neq j\leq N} \frac{1}{2} \ln |h_i-h_j|-\sum_{i=1}^N(h_i \ln h_i -h_i)\right.\\
&\hspace*{1cm}\left. +\sum_{1 \leq i\neq j\leq N}  \frac{1}{2}\ln |q^{h_j}-q^{h_i}| -\sum_{i=1}^N \sum_{j=1}^{h_i} \ln (1-q^j)\right]\exp\left[-\frac{\ln q}{6}N(N-1)(N-2)\right]\ .
\end{aligned}
\end{equation}

\section{Calculation of automodel density for the second class \`{a} la Strahov}\label{app:strahov}

Here, we give details of derivation for the solution of \eqref{eom2}. 
The first term in the kernel, namely $\coth \frac{\lambda(h-h')}{2}$ has a series representation of the form
\begin{equation}
\label{coth-expnsn}
\coth \frac{\lambda(h-h')}{2}=\frac{2}{\lambda(h-h')}+2\sum_{k=1}^{\infty}\frac{B_{2k}\lambda^{2k-1}}{(2k)!}(h-h')^{2k-1},
\end{equation}
clearly demonstrating that there is a singularity in the first term above as $h' \rightarrow h$. However, the second term is well behaved for any real $h'$. Motivated by this observation, we define the function
\begin{equation}
\label{G-defn}
\mathcal{G}(h) \equiv \int_a^b dh'\ \tilde{u}(h')\left[\coth \frac{\lambda(h-h')}{2}-\frac{2}{\lambda(h-h')}\right] .
\end{equation}
Note that in the above definition, the integral kernal has been precisely defined so as to remove any singularities. Further, since we are looking for $\tilde{u}(h)$ that are analytic within its region of support, it is expected that the function $\mathcal{G}(h)$ as defined above will be ``well-behaved". This motivates us to make the assumption that under complexification, the function will be analytic in the annulus where $a<|h|<b$. One can easily see that asymptotically, 
\begin{equation}
\mathcal{G}(h) \xrightarrow{h \to \infty} 1- \frac{2}{\lambda h} -\frac{2}{\lambda h^2}\left(k' +\frac{1}{2}\right)+O\left(\frac{1}{h^3}\right) .
\end{equation}
Note that without knowing the function $\tilde{u}(h)$ we cannot write an expression for $\mathcal{G}(h)$. However, plugging in \eqref{coth-expnsn} in \eqref{eom2}, we get,
\begin{equation}
\Xint-_{a}^{b} dh'\ \tilde{u}(h')\left[\frac{1}{h-h'}+\frac{1}{2}\sum_{k=1}^{\infty}\frac{B_{2k}\lambda^{2k}}{(2k)!}(h-h')^{2k-1}\right]=\frac{\lambda}{4}(1-a)+\frac{1}{2}\ln \frac{1-e^{-\lambda(h-a)}}{\lambda(h-a)}+\ln \frac{h-a}{\xi} .
\end{equation}
The second term appearing on the LHS of the above equation can be represented as follows:
\begin{equation}
\label{modifiedeom}
\begin{aligned}
\frac{1}{2}\sum_{k=1}^{\infty}\frac{B_{2k}\lambda^{2k}}{(2k)!}\Xint-_{a}^{b} dh'\ \tilde{u}(h')(h-h')^{2k-1}&=\frac{1}{2}\sum_{k=1}^{\infty}\frac{B_{2k}\lambda^{2k}}{(2k)!}\int_{a}^{b} dh'\ \tilde{u}(h')(h-h')^{2k-1}\\
&=\frac{1}{2}\sum_{k=1}^{\infty}\frac{B_{2k}\lambda^{2k}}{(2k)!}\sum_{n=0}^{2k-1} {2k-1 \choose n}\mathfrak{h}_{2k-n-1}(h-a)^n
\end{aligned}
\end{equation}
where 
\begin{equation}
\label{shifted-moment}
\mathfrak{h}_{n} \equiv \int_0^{b-a}\tilde{u}(h'+a)h'^n\ dh' .
\end{equation}
In writing the final line in \eqref{modifiedeom}, we first made the substitution $h' \rightarrow h'-a$ followed by a binomial expansion which brings in a second sum running from 0 to $2k-1$. We perform these operations simply for computational convenience later. Note that the quantity $\mathfrak{h}_n$ appearing in \eqref{shifted-moment} are something that we will henceforth refer to as \emph{translated moments}, are actually different from the moments of the solution $\tilde{u}(h)$.

Finally, the integral equation we require to solve takes the form
\begin{equation}
\label{utilde-eqn}
\begin{aligned}
\Xint-_{a}^{b} dh'\ \frac{\tilde{u}(h')}{h-h'}=\ln \frac{h-a}{\xi}+\frac{\lambda}{4}(1-a)&+\frac{1}{2}\ln \frac{1-e^{-\lambda(h-a)}}{\lambda(h-a)}\\
&-\frac{1}{2}\sum_{k=1}^{\infty}\frac{B_{2k}\lambda^{2k}}{(2k)!}\sum_{n=0}^{2k-1} {2k-1 \choose n}\mathfrak{h}_{2k-n-1}(h-a)^n .
\end{aligned}
\end{equation}
The resolvent corresponding to the above integral equation is given by
\begin{equation}
\label{fullresolvent}
\begin{aligned}
H(h)=\ln \frac{h}{h-a}&-\sqrt{(h-a)(h-b)}\oint_{\mathcal{C}}\frac{ds}{2\pi i}\frac{\ln \frac{s-a}{\xi}}{(s-h)\sqrt{(s-a)(s-b)}}\\
&-\sqrt{(h-a)(h-b)}\oint_{\mathcal{C}}\frac{ds}{2\pi i}\frac{\frac{\lambda}{4}(1-a)}{(s-h)\sqrt{(s-a)(s-b)}}\\
&-\sqrt{(h-a)(h-b)}\oint_{\mathcal{C}}\frac{ds}{2\pi i}\frac{\frac{1}{2}\ln \frac{1-e^{-\lambda(s-a)}}{\lambda(s-a)}}{(s-h)\sqrt{(s-a)(s-b)}}\\
&-\sqrt{(h-a)(h-b)}\oint_{\mathcal{C}}\frac{ds}{2\pi i}\frac{-\frac{1}{2}\sum_{k=1}^{\infty}\frac{B_{2k}\lambda^{2k}}{(2k)!}\sum_{n=0}^{2k-1} {2k-1 \choose n}\mathfrak{h}_{2k-n-1}(s-a)^n}{(s-h)\sqrt{(s-a)(s-b)}}
\end{aligned}
\end{equation}
where the contour $\mathcal{C}$ encloses the branch cut in the interval $[a,b]$ on the real axis (as depicted below).
\begin{figure}[H]
\begin{center}
		\includegraphics[scale=0.18]{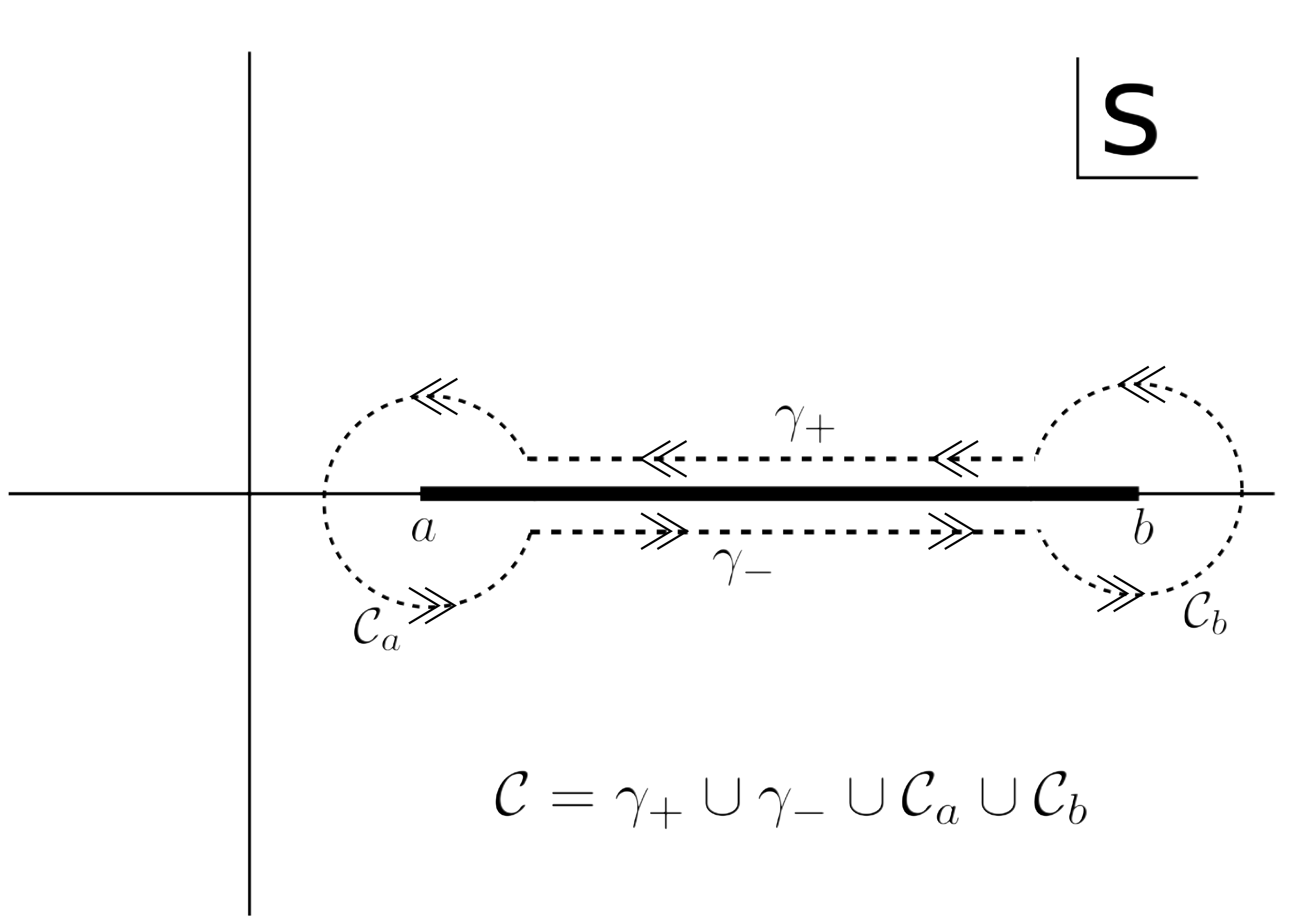}
		\label{C-contour}
		\caption{The contour $\mathcal{C}$ over which the resolvent \eqref{fullresolvent} is defined encloses the branch cut over the real interval $[a,b]$. Contour $\mathcal{C}$ is composed of four smaller compenents, namely $\gamma_{+},\gamma_{-}, \mathcal{C}_a$ and $\mathcal{C}_b$.}
	\end{center}
\end{figure}
The integral apearing in the first line above in \eqref{fullresolvent}, is in fact identical to the GWW solution and was evaluated in \cite{duttagopakumar}. Thus, using the result from \cite{duttagopakumar}, the resolvent simplifies to
\begin{equation}
H(h)=\ln \frac{h}{\xi}+\ln \frac{\sqrt{b}-\sqrt{a}}{\sqrt{b}+\sqrt{a}}+\ln \frac{h+\sqrt{ab}-\sqrt{(h-a)(h-b)}}{h-\sqrt{ab}+\sqrt{(h-a)(h-b)}} +H_{const}(h)+H_{\lambda}(h)+H_{series}(h)
\end{equation}
where the terms $H_{const}(h), H_{\lambda}(h)$ and $H_{series}(h)$ each correspond to the second, third and fourth line appearing in \eqref{fullresolvent}. 

Unlike the method employed in \cite{duttagopakumar}, we will not be blowing up the countour and extending to infinity in order to evaluate $H_{const}(h), H_{\lambda}(h)$ and $H_{series}(h)$ but rather compute it directly through an appropriate choice of parametrization of the integration variable $s$. See \ref{resolvent-calc} for details of the computation. After evaluating all the contour integrals, the final expression of resolvent is given by
\begin{equation} 
\label{resolvent-exact}
\begin{aligned}
 H(h)&=\ln \frac{h}{\xi}+\ln \frac{\sqrt{b}-\sqrt{a}}{\sqrt{b}+\sqrt{a}}+\ln \frac{h+\sqrt{ab}-\sqrt{(h-a)(h-b)}}{h-\sqrt{ab}+\sqrt{(h-a)(h-b)}}+\frac{\lambda}{4}(1-a)+\sum_{k=1}^{\infty}A_{k}^{(1)}\sqrt{\frac{(h-a)}{(h-b)}}\\&\hspace*{2cm}+\sum_{k=1}^{\infty}A_{k}^{(2)}\frac{_2F_1(1,\frac{1}{2}+k,2+k,\frac{a-b}{a-h})}{\sqrt{(h-a)(h-b)}}+\sum_{k=1}^{\infty}\sum_{n=1}^{2k-1}A_{k,n}^{(1)}\sqrt{\frac{(h-a)}{(h-b)}}\\&\hspace*{2cm}+\sum_{k=1}^{\infty}\sum_{n=1}^{2k-1}A_{k,n}^{(2)}\frac{_2F_1(1,\frac{1}{2}+n,2+n,\frac{a-b}{a-h})}{\sqrt{(h-a)(h-b)}} ,       
\end{aligned}
\end{equation}
where
\begin{eqnarray}\label{eq:Aks}
A_{k}^{(1)}&=&\frac{1}{4\sqrt{\pi}}\frac{\lambda^{k}B_{k}}{k.k!}(b-a)^{k}\Gamma(k+\frac{1}{2})\frac{2}{\Gamma(k+1)}\ ,\nonumber\\
A_{k}^{(2)}&=&-\frac{1}{4\sqrt{\pi}}\frac{\lambda^{k}B_{k}}{k.k!}(b-a)^{k+1}\Gamma(k+\frac{1}{2})\ ,\nonumber\\
A_{k,n}^{(1)}&=&-\frac{1}{4\sqrt{\pi}}\frac{\lambda^{2k}B_{2k}}{2k!}{2k-1 \choose n}\mathfrak{h}_{2k-n-1}(b-a)^{n}\Gamma(n+\frac{1}{2})\frac{2}{\Gamma(n+1)}\nonumber\\
A_{k,n}^{(2)}&=&\frac{1}{4\sqrt{\pi}}\frac{\lambda^{2k}B_{2k}}{2k!}{2k-1 \choose n}\mathfrak{h}_{2k-n-1}(b-a)^{n+1}\Gamma(n+\frac{1}{2}) .
\end{eqnarray}

\subsection{{Computing $H_{const}(h), H_{\lambda}(h)$ and $H_{series}(h)$}}
\label{resolvent-calc}
In order to perform the contour integrals in \eqref{fullresolvent}, one needs to reparametrize the integration variable $s$ as $s=a+xe^{i\epsilon}$ along $\gamma_{+}$ and $s=a+xe^{i(2\pi -\epsilon)}$ along $\gamma_{-}$. It can be further demonstrated that the contribution to this integral coming from the parts $\mathcal{C}_a$ and $\mathcal{C}_b$ i.e. two infinitesimal circles centered around $a$ and $b$ vanishes in the limit when the radii of the circles are taken to zero.

Carrying out the above procedure, we get
\begin{equation}
H_{const}(h)=-\sqrt{(h-a)(h-b)}\oint_{\mathcal{C}}\frac{ds}{2\pi i}\frac{\frac{\lambda}{4}(1-a)}{(s-h)\sqrt{(s-a)(s-b)}}=\frac{\lambda}{4}(1-a) .
\end{equation}
For the next term, we get,
\begin{equation}
\begin{aligned}
H_{\lambda}(h)&=-\sqrt{(h-a)(h-b)}\oint_{\mathcal{C}}\frac{ds}{2\pi i}\frac{\frac{1}{2}\ln \frac{1-e^{-\lambda(s-a)}}{\lambda(s-a)}}{(s-h)\sqrt{(s-a)(s-b)}}\\
&=-\frac{\sqrt{(h-a)(h-b)}}{2}\sum_{k=1}^{\infty}\frac{\lambda^{k}B_{k}}{(k).(k)!}\oint_{\mathcal{C}}\frac{ds}{2\pi i}\frac{(s-a)^{k}}{(s-h)\sqrt{(s-a)(s-b)}}\ .
\end{aligned}
\end{equation}
In writing the above line, we have used the expansion
\begin{equation}
\ln \frac{1-e^{-\lambda(s-a)}}{\lambda(s-a)}=\sum_{n=1}^{\infty}\frac{\lambda^{n}B_{n}}{(n).(n)!}(s-a)^{n}\ .
\end{equation}
Simplifying the above integrals, we get eventually
\begin{equation}
\begin{aligned}
H_{\lambda}(h)&=-\frac{\sqrt{(h-a)(h-b)}}{2\pi }\sum_{k=1}^{\infty}\frac{\lambda^{k}B_{k}}{(k).(k)!}\int_{0}^{b-a}dx \frac{x^{k}}{(x+a-h)\sqrt{x(b-a-x)}}\\
&=-\frac{\sqrt{(h-a)(h-b)}}{4\sqrt{\pi}}\sum_{k=1}^{\infty}\frac{\lambda^{k}(b-a)^{k}B_{k}}{(b-h)(k).(k)!}\Gamma(k+\frac{1}{2})\\&\hspace*{2cm}\left[\frac{2}{\Gamma(k+1)}+ \big(\frac{b-a}{a-h}\big)._2F_1(1,\frac{1}{2}+k,2+k,\frac{a-b}{a-h})\right] .
\end{aligned}
\end{equation}
A similar exercise on $H_{series}(h)$ yields:
\begin{equation}
\begin{aligned}
H_{series}(h)&=-\sqrt{(h-a)(h-b)}\oint_{\mathcal{C}}\frac{ds}{2\pi i}\frac{-\frac{1}{2}\sum_{k=1}^{\infty}\frac{B_{2k}\lambda^{2k}}{(2k)!}\sum_{n=0}^{2k-1} {2k-1 \choose n}\mathfrak{h}_{2k-n-1}(s-a)^n}{(s-h)\sqrt{(s-a)(s-b)}}\\
&=\frac{\sqrt{(h-a)(h-b)}}{2}\cdot \sum_{k=1}^{\infty}\frac{B_{2k}\lambda^{2k}}{(2k)!}\sum_{n=0}^{2k-1} {2k-1 \choose n}\mathfrak{h}_{2k-n-1}\times\\&\hspace*{5cm} \oint \frac{ds}{2\pi i}
\frac{(s-a)^n}{(s-h)\sqrt{(s-a)(s-b)}}\\
&=\frac{\sqrt{(h-a)(h-b)}}{4\sqrt{\pi}}\frac{1}{(b-h)}\cdot \sum_{k=1}^{\infty}\frac{B_{2k}\lambda^{2k}}{(2k)!}\sum_{n=0}^{2k-1} {2k-1 \choose n}\mathfrak{h}_{2k-n-1}(b-a)^n\times\\
&\hspace*{2cm}\Gamma\left(n+\frac{1}{2}\right)\left[\frac{2}{\Gamma(n+1)}-\frac{a-b}{a-h}\cdot {_2F_1\left[1,n+\frac{1}{2};n+2; \frac{a-b}{a-h}\right]}\right] .
\end{aligned}
\end{equation}

\section{Droplet quantisation}\label{app:hi}

The droplet picture, discussed in section \ref{sec:umm-droplet}, has no dynamics.  Different large $N$ phases, obtained by solving the saddle point equation, are possible minimum free energy configurations of the model. Hence the corresponding droplets are static - the shapes are not changing with time. To incorporate dynamics into the picture, we follow an ad-hoc way : identify the large $N$ droplets with Thomas-Fermi distribution at zero temperature and obtain the single particle Hamiltonian by comparing the two. In this way it is possible to incorporate time in our system.

Comparing the phase space distribution function (\ref{eq:phasespacedistri}) with Thomas-Fermi distribution function, we get
\ben\label{eq:singleh}
\mathfrak{h}(h,\q)= {h^2\over 2} - S(\q)h + {g(\q)\over 2} + \mu, \quad 
\where \quad g(\q) = h_+(\q) h_-(\q).
\een
The above Hamiltonian is not in diagonal form. Defining a new variable 
\ben
\bar h = h-S(\theta).
\een
the single particle Hamiltonian density is given by
\ben\label{eq:singleh2}
\mathfrak{h}(\bar h,\theta) = \frac{\bar h^2}2 -\frac{\pi^2}2 \rho(\theta)^2
\een
and the droplet boundary is given by
\ben\label{eq:boundaryeqn2}
\bar h_{\pm}(\theta) = \pm \pi \rho(\theta).
\een
The phase space Hamiltonian can be calculated by integrating (\ref{eq:singleh2}) over $\bar h$ from $\bar h_{-}(\theta)$ to $\bar h_{+}(\theta)$ and is given by
\ben\label{eq:Hh1}
H_h = \frac1{3\pi \hbar}\int d\theta \ \pi^3 \rho(\theta)^3
\een
where $\hbar =1/N$ is the minimum accessible area in phase space. The phase space boundary is doubly degenerate, i.e. for a given $\theta$ there are two boundaries, $\bar h_{\pm}$ corresponding to two signs on the right hand side of (\ref{eq:boundaryeqn2}). Hence, we write the Hamiltonian (\ref{eq:Hh1}) over the two segments of droplets \cite{Maoz:2005nk}
\ben\label{eq:Hhtwoboundary}
H_h = \frac1{6\pi \hbar}\left[\int_+ d\theta \ \bar h^3_+(t,\theta) - \int_- d\theta \ \bar h^3_-(t,\theta)  \right].
\een
From the single particle Hamiltonian (\ref{eq:singleh2}), we find that the equations of motion satisfied by the droplet boundaries are given by\footnote{Equations of motion from (\ref{eq:singleh2}) are given by %
\ben\label{eq:eom}
\dot{\bar h}(t) = -\frac{\partial \mathfrak h}{\partial \theta} = \pi^2 \rho(\theta) \rho'(\theta), \qquad \dot \theta(t) = \frac{\partial \mathfrak h}{\partial \bar h} = \bar h(t)
\een
Substituting the boundary relations (\ref{eq:boundaryeqn2}) in these equations, one finds (\ref{eq:boundeq}).}
\ben\label{eq:boundeq}
\dot{\bar h}_{\pm}(t,\theta) = \bar h_\pm(t,\theta)\bar h_\pm'(t,\theta)
\een
where the dot and prime denote derivative with respect to $t$ and $\theta$ respectively.

In order to quantise the droplets, we find a simplectic form such that the Hamilton's equation
\be\label{eq:Hameq2}
\dot {\bar h}_{\pm}(t,\theta) = \{\bar h_\pm(t,\theta),H_h\}
\ee
would coincide with (\ref{eq:boundeq}) \cite{Grant:2005qc,Maoz:2005nk}. To achieve this, we introduce Poisson brackets between $\bar h_{\pm}(t,\theta)$ and $\bar h_{\pm}(t,\theta')$
\ben\label{eq:poisson}
\{\bar h_\pm(t,\theta),\bar h_\pm(t,\theta')\} = \pm {\pi \hbar} \delta'(\theta-\theta')\quad \text{and} \quad \{\bar h_+(t,\theta),\bar h_-(t,\theta')\} =0.
\een
It is easy to check that using these Poisson brackets, equation (\ref{eq:Hameq2}) boils down to (\ref{eq:boundeq}).

The circular droplet ($\rho(\theta) =\frac1{2\pi}$) corresponds to $\bar h_\pm(t,\theta)=\pm \frac12$ and satisfy the classical equations of motion (\ref{eq:boundeq}). We consider quantum fluctuations about this classical circular droplet
\ben\label{eq:hpmode}
\bar h_{\pm}(t,\theta) = \pm\frac12+ \hbar\ \tilde h_\pm(t,\theta).
\een
We also assume that the fluctuations preserve the total area of the droplet. This implies 
\ben\label{eq:tildehcons}
\int_{-\pi}^{\pi} d\theta (\tilde h_+(t,\theta) - \tilde h_-(t,\theta))=0.
\een

To quantise the above classical system, we promote the Poisson brackets (\ref{eq:poisson}) to commutation relations
\be\label{eq:commutation}
\left[ \tilde h_{\pm}(t,\theta), \tilde h_{\pm}(t,\theta')\right] = \pm \pi i \delta'(\theta-\theta')\quad \text{and}\quad [\tilde h_{+}(t,\theta),\tilde h_-(t,\theta')]=0.
\ee
We decompose $\tilde{h}_\pm(t,\theta)$ into Fourier modes
\ben\label{eq:hpmode}
\tilde h_{+}(t,\theta) = \sum_{n=-\infty}^\infty a_{-n}(t) e^{i n \theta}, \quad \tand \quad \tilde h_{-}(t,\theta) =  - \sum_{n=-\infty}^\infty b_{n}(t) e^{i n \theta}.
\een
From the constraint equation (\ref{eq:tildehcons}), we see that the zero-modes $a_0$ and $b_0$ are equal up to a sign
\ben
a_0=-b_0=\pi_0.
\een
It follows from equation (\ref{eq:commutation}) that the Fourier modes $a_n$ and $b_n$ satisfy a $U(1)$ \emph{Kac-Moody} algebra
\ben \label{eq:U1KM}
[a_m(t),a_n(t)]={\frac{1}{2}}m\delta_{m+n},\quad [b_m(t),b_n(t)]={\frac{1}{2}}m\delta_{m+n}, \quad \tand \quad [a_m(t),b_n(t)]=0.
\een
Since $\tilde h_\pm(t,\theta)$ are real, we have $a_{-n}=a_n^\dagger$ and $b_{-n}=b_n^\dagger$. The Hamiltonian (\ref{eq:Hhtwoboundary}) in terms of these modes is given by (up to over all constant factors)
\ben
\tilde H = H_+ + H_-
\een
where
\ben
\begin{split}
H_{+} = & \frac{\hbar}{2}a_{0}^2 + \frac{\hbar^2}{3}a_{0}^3 - \frac{\hbar^2}{24}a_{0} + \hbar (1+2\hbar a_0)\sum_{n>0} a_n^\dagger a_n \\
 &+ \hbar^2\sum_{m,n>0} \lb a_{m+n}^\dagger a_m a_n+ h.c. \rb
\end{split}
\een
and $H_-$ has a similar expression in terms of modes $b_n$'s.\\
Since $\pi_0$ commutes with all the $a_n, b_n$ and hence with $\tilde H$, application of $a_n$s and $b_n$s cannot change the eigenvalue of $\pi_0$. Therefore, the Hilbert space is constructed upon a one parameter family of vacua $\ket{s,s}\equiv \ket s$, where
\ben
\label{eq:KMprimary}
\begin{split}
   & a_n\ket{s} = 0, \quad b_n\ket{s} = 0 \quad \for\  n>0\\
\tand \quad & a_0\ket{s} = - b_0\ket{s} =\pi_0\ket{s} = s \ket{s}.
\end{split}
\een
Starting with the primary $\ket{s}$ we can now construct a Hilbert space $\cH$ which is an $s$ charged module. Let us denote by $\cH_+$ and $\cH_-$ as Hilbert spaces associated with $a$ and $b$ sectors respectively. The commutativity of $a$ and $b$ operators implies that the full Hilbert space is factorizable into $\cH_+$ and $\cH_-$ i.e. $\cH=\cH_+ \otimes \cH_-$.

A generic excited state in $\cH$ is given by
\ben\label{eq:basisstatek}
\ket{\vec k,\vec l} = \prod_{n,m=1}^{\infty} \lb a^\dagger_n\rb^{k_n}\lb b^\dagger_m\rb^{l_m}\ket{s}.
\een
The vectors $\vec k$ and $\vec l$ correspond to excitations in the upper and lower Fermi surfaces. Since $a_n$ and $b_n$ commute, a generic excitation $\ket{\vec k, \vec l} \in \cH$ can be written as a direct product of states belonging to the two sectors i.e. $\ket{\vec k, \vec l}=\ket{\vec k} \otimes \ket{\vec l}$.

The excited states in $\cH_+$ are orthogonal with the normalization
\ben\label{eq:knormalization}
\langle {\vec{k'}}|{\vec k}\rangle =z_{\vec k}\delta_{\vec k \vec{k'}}\ \ \text{where}\ z_{\vec{k}}=\prod_j k_j! {\left(\frac{j}{2}\right)}^{k_j}
\een
and has $\pi_0$ eigenvalue $s$. They also satisfy the completeness relation
\ben \label{eq:completeness}
\sum_{\vec k}\frac{1}{z_{\vec k}} \ket{\vec k}\bra{\vec k} = {\mathbb{I}}_{\cH_+}
\een
and hence form a basis in $\cH_+$. The particle like excited states $\ket{\vec k}$ in either sectors are eigenstates of the free Hamiltonian of the corresponding sectors
\ben
H^\pm_{\text{free}}\ket{\vec k} = \hbar(1 \pm 2s\hbar)\lb\sum_{n=1}^\infty n k_n\rb \ket{\vec k},
\een
but not an eigenstate of the full Hamiltonian. The interaction Hamiltonian is such that its action on an excited state doesn't change the level of the state. Explicitly stated, the interaction Hamiltonian takes a state $\ket{\vec k}$ to $\ket{\vec{k'}}$ satisfying $\sum_n n k_n =\sum_n n{k'}_n$. The expectation value of $\bar h_+(t,\theta)$ operator in $\ket{\vec k}$ state is $(1/2+\hbar s)$. The quantum dispersion, $\Delta \bar{h}_+$ in $\ket{\vec k}$ state is $O(\hbar)$ and hence a generic $\ket{\vec k}$ state can be interpreted as quantum fluctuations over the ground state.

A generic coherent state in $\cH_+$ can be defined as
\ben\label{eq:coherentstate}
\ket{\t_+} = \exp\lb {\sum_{n=1}^\infty \frac{2\t_{n}^+ a_n^\dagger}{n\hbar}}\rb \ket{s}.
\een
The state $\ket{\t_+}$ is not normalised. The normalization is given by 
\ben
\braket{\tau_+^a}{\tau^b_+} = \exp\lb\sum_n \frac{4\t_{n+}^a\t_{n+}^b}{n\hbar^2}\rb.
\een
The coherent state $\ket{\t_+} $ is an eigenstate of $a_n$ ($\forall\, n>0$) with eigenvalue $\t_{n}^+/\hbar$. A coherent state $\ket{\t_+}$ can be expanded in $\ket{\vec k}$ basis in the following way
\ben
\ket {\t_+} = \sum_{\vec k} \frac{\t^+_{\vec k}}{z_{\vec k}}\ket{\vec k}, \quad \where \quad \t^+_{\vec k} = \prod_m \lb\frac{\t^+_{m}}{\hbar}\rb^{k_m}.
\een
The expectation value of $ \bar{h}_+$ in a coherent state $\ket {\t_+}$ is given by
\ben
\omega_{\t_+}(z)= \frac{ \bra{\t_+}\frac{\bar h_+(z)}{\pi} \ket{\t_+}}{\braket{\t_+}{\t_+}} =\frac{1}{2\pi} + \frac{s\hbar}{\pi}+\frac{1}{\pi}\sum_{n>0}\t^+_{n}\lb z^n +\frac{1}{z^n}\rb.
\een
Value of $\omega_{\t_+}(z)$ on a unit circle ($|z|=1$) in the complex $z$ plane is given by 
\ben\label{eq:coherentprofile}
\begin{split}
\omega_{\t_+}(\theta) \equiv \omega_{\t_+}(z=e^{i\theta}) & = \frac{1}{2\pi}  + \frac{s\hbar}{\pi} + \tilde{\omega}_{\t_+}(\theta)\\
\text{where} \quad \tilde{\omega}_{\t_+}(\theta) &=\frac{2}{\pi} 
\sum_{n>0}\t^+_{n} \cos n\theta.
\end{split}
\een
Since the quantum dispersion of $\bar{h}_+$ in a coherent state is zero, we call such states \emph{classical}.

\bibliographystyle{hieeetr}
\bibliography{PlancherelNotes.bib}

\begin{thebibliography}{10}

\bibitem{Chattopadhyay:2019pkl}
A.~Chattopadhyay, S.~Dutta, D.~Mukherjee, and Neetu, ``{Quantum Mechanics of
  Plancherel Growth},'' 2019, 1909.06797.

\bibitem{Eynard:2008mt}
B.~Eynard, ``{All orders asymptotic expansion of large partitions},'' {\em J.
  Stat. Mech.}, vol.~0807, p.~P07023, 2008, 0804.0381.

\bibitem{VerKer77}
A.~Vershik and S.~Kerov, ``{Asymptotics of the Plancherel measure of the
  symmetric group and the limiting form of Young tableaux},'' {\em Dokl. Akad.
  Nauk}, vol.~233, pp.~1024--1027, 1977.

\bibitem{LogShe}
B.~Logan and L.~Shepp, ``{A variational problem for random Young tableaux},''
  {\em Advances in Mathematics}, vol.~26, pp.~206--222, 1977.

\bibitem{Kerov1}
{S. V. Kerov}, ``{A Differential Model Of Growth Of Young Diagrams},'' {\em
  Proceedings of St.Petersburg Mathematical Society}, 1996.

\bibitem{duttagopakumar}
S.~Dutta and R.~Gopakumar, ``{Free fermions and thermal $AdS/CFT$},'' {\em
  JHEP}, vol.~03, p.~011, 2008, 0711.0133.

\bibitem{riemannzero}
P.~Dutta and S.~Dutta, ``{Phase Space Distribution of Riemann Zeros},'' {\em J.
  Math. Phys.}, vol.~58, no.~5, p.~053504, 2017, 1610.07743.

\bibitem{Chattopadhyay}
A.~Chattopadhyay, P.~Dutta, and S.~Dutta, ``{Emergent Phase Space Description
  of Unitary Matrix Model},'' {\em JHEP}, vol.~11, p.~186, 2017, 1708.03298.

\bibitem{Strahov07}
E.~{Strahov}, ``{A differential Model for the Deformation of the Plancherel
  Growth Process},'' {\em arXiv e-prints}, p.~arXiv:0706.3292, Jun 2007,
  0706.3292.

\bibitem{Caporaso:2006gk}
N.~Caporaso, L.~Griguolo, M.~Marino, S.~Pasquetti, and D.~Seminara, ``{Phase
  transitions, double-scaling limit, and topological strings},'' {\em Phys.
  Rev. D}, vol.~75, p.~046004, 2007, hep-th/0606120.

\bibitem{Nekrasov:2003rj}
N.~Nekrasov and A.~Okounkov, ``{Seiberg-Witten theory and random partitions},''
  {\em Prog. Math.}, vol.~244, pp.~525--596, 2006, hep-th/0306238.

\bibitem{Nekrasov:2002qd}
N.~A. Nekrasov, ``{Seiberg-Witten prepotential from instanton counting},'' {\em
  Adv. Theor. Math. Phys.}, vol.~7, no.~5, pp.~831--864, 2003, hep-th/0206161.

\bibitem{Marino:2006hs}
M.~Marino, ``{Open string amplitudes and large order behavior in topological
  string theory},'' {\em JHEP}, vol.~03, p.~060, 2008, hep-th/0612127.

\bibitem{okounkov2006uses}
A.~Okounkov, ``The uses of random partitions,'' in {\em XIVth International
  Congress on Mathematical Physics}, pp.~379--403, World Scientific, 2006.

\bibitem{Chattopadhyay:2020rle}
A.~Chattopadhyay, S.~Dutta, D.~Mukherjee, and Neetu, ``{From 2d Droplets to 2d
  Yang-Mills},'' 10 2020, 2010.11923.

\bibitem{fulton1991representation}
W.~Fulton and J.~Harris, {\em Representation Theory: A First Course}.
\newblock Graduate texts in mathematics, Springer, 1991.

\bibitem{Kerov1993AQO}
S.~Kerov, ``A q-analog of the hook walk algorithm for random young tableaux,''
  {\em Journal of Algebraic Combinatorics}, vol.~2, pp.~383--396, 1993.

\bibitem{2010arXiv1001.2180F}
V.~{Feray} and P.-L. {M{\'e}liot}, ``{Asymptotics of q-Plancherel measures},''
  {\em arXiv e-prints}, p.~arXiv:1001.2180, Jan. 2010.

\bibitem{Migdal:1984gj}
A.~A. Migdal, ``{Loop Equations and 1/N Expansion},'' {\em Phys. Rept.},
  vol.~102, pp.~199--290, 1983.

\bibitem{1998math.10105B}
J.~{Baik}, P.~{Deift}, and K.~{Johansson}, ``{On the Distribution of the Length
  of the Longest Increasing Subsequence of Random Permutations},'' {\em arXiv
  Mathematics e-prints}, Oct 1998, math/9810105.

\bibitem{duttadutta}
P.~Dutta and S.~Dutta, ``{Phase Space Distribution for Two-Gap Solution in
  Unitary Matrix Model},'' {\em JHEP}, vol.~04, p.~104, 2016, 1510.03444.

\bibitem{BIPZ}
E.~Brezin, C.~Itzykson, G.~Parisi, and J.~B. Zuber, ``{Planar Diagrams},'' {\em
  Commun. Math. Phys.}, vol.~59, p.~35, 1978.

\bibitem{douglas2}
M.~R. Douglas, ``{Conformal field theory techniques in large $N$ Yang-Mills
  theory},'' in {\em {NATO Advanced Research Workshop on New Developments in
  String Theory, Conformal Models and Topological Field Theory Cargese, France,
  May 12-21, 1993}}, 1993, hep-th/9311130.

\bibitem{Jevicki:1993qn}
A.~Jevicki, ``{Development in 2-d string theory},'' in {\em {Workshop on String
  Theory, Gauge Theory and Quantum Gravity}}, 9 1993, hep-th/9309115.

\bibitem{Maoz:2005nk}
L.~Maoz and V.~S. Rychkov, ``{Geometry quantization from supergravity: The Case
  of `Bubbling AdS'},'' {\em JHEP}, vol.~08, p.~096, 2005, hep-th/0508059.

\bibitem{Grant:2005qc}
L.~Grant, L.~Maoz, J.~Marsano, K.~Papadodimas, and V.~S. Rychkov,
  ``{Minisuperspace quantization of `Bubbling AdS' and free fermion
  droplets},'' {\em JHEP}, vol.~08, p.~025, 2005, hep-th/0505079.

\end{thebibliography}

\end{document}